\documentclass{aa}  
\usepackage{graphicx}
\usepackage{txfonts}
\usepackage{gensymb}
\usepackage[colorlinks=true, citecolor=blue]{hyperref}
\usepackage{multirow}
\usepackage{placeins}
\usepackage{soul}

\def \arcsec{\text{$^{\prime\prime}$}}

\begin{document} 

   \title{Two-component large-scale radio emission in Abell 2244}

    \author{M.~Cianfaglione \inst{1,2}
        \and F.~De~Gasperin \inst{1}
        \and V.~Cuciti \inst{2,1}
        \and M.~Balboni \inst{2,3}
        \and R. J. van Weeren \inst{4}
        \and C. Groeneveld \inst{1}
        \and J. M. Boxelaar \inst{1,2}
        \and M. Della Chiesa \inst{1,2}
        \and A. Bonafede \inst{2,1}
        \and G. Di Gennaro \inst{1}
        \and F. Gastaldello \inst{3} 
        \and G. Brunetti \inst{1}
          }

    \institute{INAF - Istituto di Radioastronomia, via P. Gobetti 101, Bologna, Italy \\
            \email{m.cianfaglione@ira.inaf.it}
        \and Dipartimento di Fisica ed Astronomia, Università di Bologna, Via Gobetti 93/2, 40129 Bologna, Italy
        \and INAF - IASF Milano, via A. Corti 12, 20133 Milano, Italy 
        \and Leiden Observatory, Leiden University, PO Box 9513, 2300 RA Leiden, The Netherlands 
             }

   \date{Received ???; accepted ???}

  \abstract
    {In recent years, clusters have been observed that host multi-component haloes, both in non-merging and merging systems. The existence of these multi-component haloes suggests that there is no clear distinction between the single components.}
    {Abell 2244 is an intermediate-mass cluster that hosts a double component diffuse radio emission. The aim of this paper is to carry out a in-depth study of the diffuse radio emission to constrain its origin and characterize its main radio properties.}
    {In this work we present LOFAR HBA, MeerKAT UHF, and L-band observations of the cluster Abell 2244. We investigated the nature of the diffuse radio emission, combining high sensitivity radio data with \textit{XMM-Newton} deep X-ray observations. We also used mock LOFAR observations to investigate contamination of the emission from faint radio sources.}
    {We find an integrated spectral index of $\alpha^{1279}_{144} = 0.9 \pm 0.1$ for both components, where only the radio halo shows spectral steepening at higher frequencies. These values are comparable with the spectral indices observed in disturbed massive clusters. The outer component does not follow the same radio X-ray correlation as the radio halo, which suggests a different physical origin.}
    {By analysing the physical and morphological properties of the diffuse emission, we find that the characteristics of the outer component of the emission are intermediate between those of radio haloes and of known megahaloes. Hence, we speculate that the source is either a morphologically disturbed radio halo, caused by a minor merger interaction, or a megahalo but we cannot reach a final classification. From the mock observations, we find that it is unlikely that the emission is caused by faint sources at low resolutions.}


   \maketitle

\section{Introduction}
    High sensitivity radio observations are revealing an increasing number of radio halo (RH) and mini-halo (mH) sources in galaxy clusters, alongside other kinds of diffuse radio sources \citep{vanWeeren2019}.
    The existence of these sources shows that magnetic fields and relativistic cosmic rays are present in the intra-cluster medium (ICM).
    The interaction between relativistic cosmic ray electrons (CRe) and the magnetic field results in synchrotron emission from these particles.
    If there is no continuous injection of particles, CRe lose energy via synchrotron emission and inverse Compton processes.
    Because of these losses, CRe should lose their energy before diffusing over megaparsec scales, hence a mechanism of in-situ particle acceleration or injection must be invoked to explain the existence of diffuse radio emission \citep{Jaffe1977, Brunetti2014}.
    \begin{figure*}[!hbt]
        \centering
        \includegraphics[width = 0.6\textwidth]{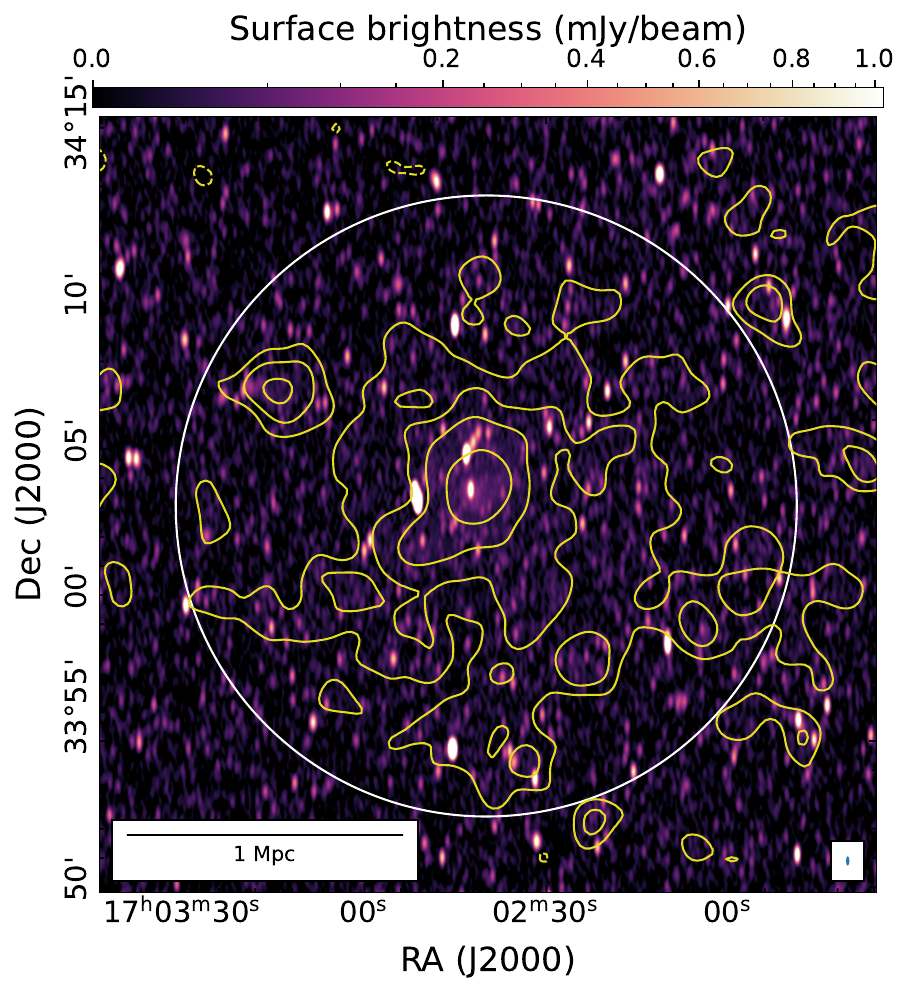}
        \includegraphics[width = 0.42\textwidth]{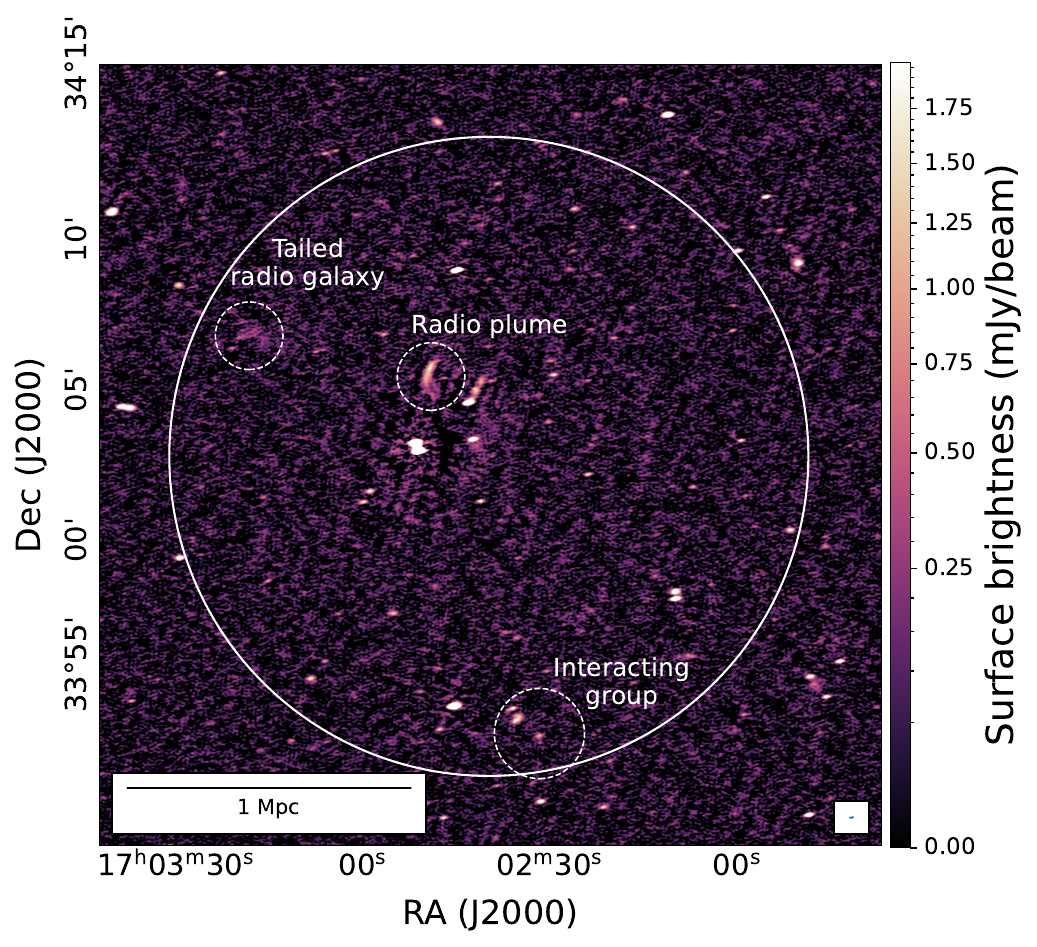}
        \includegraphics[width = 0.42\textwidth]{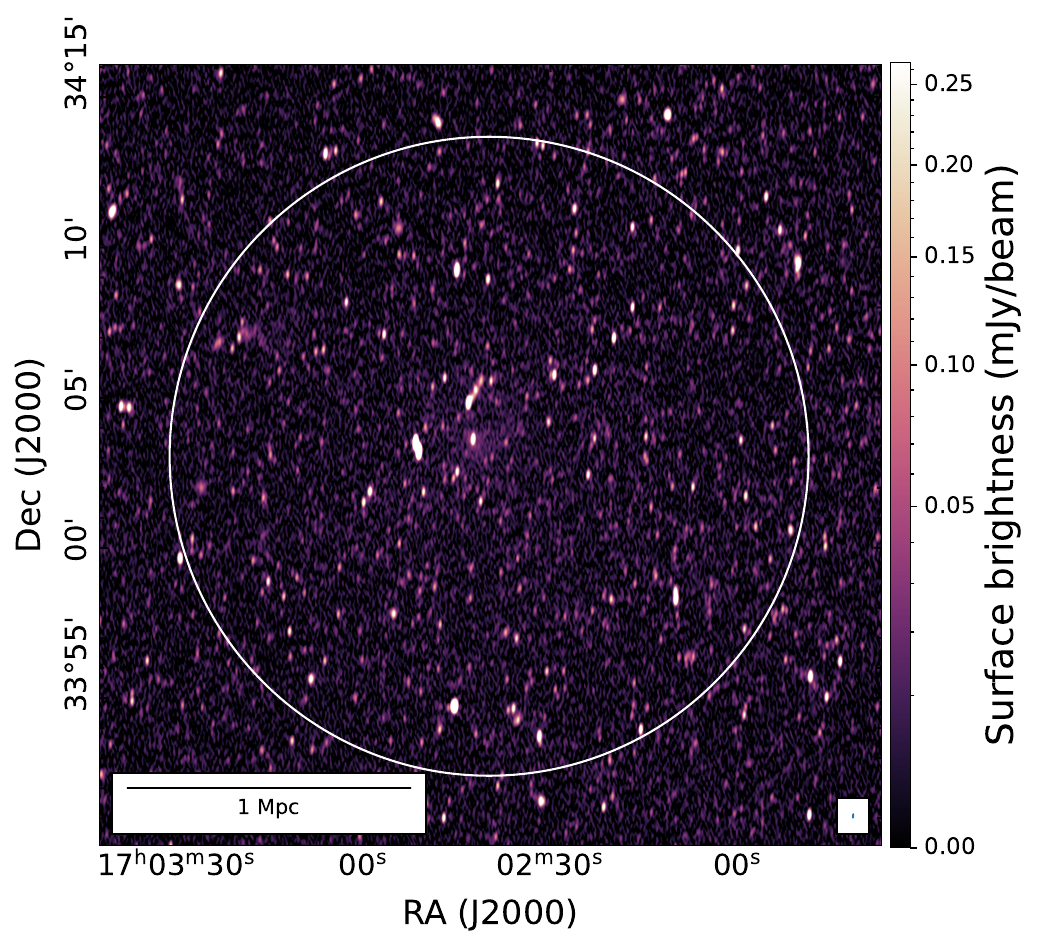}
        \caption{\textit{Top:} High-resolution ($\approx 10\arcsec$) MeerKAT UHF band image with the contours of the diffuse radio emission observed at low resolutions ($\approx 60\arcsec$) overlaid. The noise level is $\sigma_{rms} = 10 \ \mathrm{\mu Jy/beam}$. \textit{Bottom left:} High-resolution ($\approx 6\arcsec$) LOFAR HBA image. The dashed circles highlight the peculiar sources. The rms is $\sigma_{rms} = 120 \ \mathrm{\mu Jy/beam}$. \textit{Bottom right:} High-resolution ($\approx 8\arcsec$) MeerKAT L-band image. The noise is $\sigma_{rms} = 8 \ \mathrm{\mu Jy/beam}$. All the images are made with an inner uv-cut of $80\lambda$. The white circle is $r_{500}$.}
        \label{fig:sources_imgs}
        \vspace{-3mm}
    \end{figure*}
    One of the main acceleration mechanisms in clusters is related to ICM turbulence, which can be injected from the sloshing in the core \citep{ZuHone2013} or injected by merger events between clusters or groups \citep{Petrosian2001, Brunetti2001}.
    In agreement with this view, RHs, which span spatial scales in the range $0.5-1~\mathrm{r_{500}}$, defined as the radius at which the average density is 500 times the critical density of the Universe at the source redshift, are usually found in disturbed galaxy clusters \citep{Cassano2010b, Cassano2013, Cuciti2021, Cuciti2023} and are co-spatial with the X-ray emission from the ICM \citep[e.g.][]{Balboni2024}.
    Mini-haloes, instead, are found in the cores of relaxed clusters, with sizes up to $0.2~\mathrm{r_{500}}$ \citep[e.g.][]{Gitti2002, Gitti2015, Giacintucci2017}, where the sloshing is related to the core sloshing in the cluster potential well.
    Due to the low acceleration efficiency of stochastic processes, the presence of synchrotron-emitting CRe requires the existence of fossil CRe in the ICM \citep{Vazza2024}.
    The average integrated spectral index of RHs is $\alpha = 1.3$ \citep[$S_\nu \propto \nu^{-\alpha}$;][]{Feretti2012}, with a population of ultra-steep spectrum ($\alpha \gtrsim 1.6$) sources in low-mass clusters or when minor mergers take place \citep{Cuciti2021}, which is predicted by theory \citep{Cassano2010}.
    Mini-haloes have steep spectra with values similar to RHs.
    The azimuthally averaged surface brightness radial profiles of both RHs and mHs follow an exponential profile, which can be integrated to analytically compute their fluxes \citep{Murgia2009}. \par
    Recent observations showed that RHs can be present in relaxed galaxy clusters \citep{Savini2019} and can co-exist with mHs \citep{Biava2021, Biava2024}, which can be explained in cases of off-axis merger events that do not disrupt the relaxed core.
    These events can inject enough turbulence to re-accelerate CRe on larger scales \citep[e.g.][]{Biava2021, vanWeeren2024, Riseley2024, Lusetti2024}, where the outer component should have a steeper spectrum if there is less turbulent energy available outside of the central region \citep{Cassano2006, Brunetti2008}. \par
    Radio haloes and mHs show the presence of a point-to-point correlation with the X-ray surface brightness of the X-ray emission from the ICM.
    This relation is based on the similar morphologies between the X-ray and radio diffuse emissions, which can indicate a relation between the energy of the ICM particles and of CRe \citep{Govoni2001}.
    Different works in the past years have shown that the correlation for RHs is sub-linear \citep[e.g.][]{Giacintucci2005, Hoang2021, Balboni2024} and super-linear for mHs \citep[e.g.][]{Ignesti2020}. \par
    Recently, a new type of low brightness diffuse radio emission has been found in four massive ($M > 6 \times 10^{14}~\mathrm{M_\odot}$) disturbed galaxy clusters of the LOw Frequency ARray \citep[LOFAR; ][]{vanHaarlem2013} Two Metre Sky Survey (LoTSS) \textit{Planck} cluster survey \citep{Shimwell2017, Shimwell2019, Shimwell2022, Botteon2022a}, as reported in \cite{Cuciti2022}.
    These `megahaloes' (megaHs) show a characteristic radial surface brightness profile, as it appears as an outer non-exponential component around the RH.
    These sources have largest linear scales of a few megaparsecs and extend up to and beyond $r_{500}$.
    In their work, \cite{Cuciti2022} measured the spectral index between $50~\mathrm{MHz}$ and $144~\mathrm{MHz}$ for two of the four megaHs detected, and obtained spectral index values of $\alpha \approx 1.6$, making them ultra-steep spectrum sources. 
    This could be explained by the presence of low energy turbulence in the clusters outskirts.
    More recently, \cite{Rajpurohit2025} suggest that apparent megaH-like large-scale emission could be produced, at least partially, by a blend of residual emission from the incomplete subtraction of complex discrete radio sources and of faint point radio sources, which appear as diffuse emission at low resolutions. 
    The authors suggest that, to mitigate this issue, high-sensitivity observations, such as the ones provided by MeerKAT \citep{Jonas2016}, are needed to properly subtract the compact sources. \par
    Overall, high-quality spectral index analyses of complex diffuse radio sources, which are needed to constrain the acceleration mechanisms and understand the origin of these sources, are still missing.
    The study and characterization of the different types of diffuse halo sources in clusters and how they differ from each other will allow us to better understand the particle acceleration mechanisms in these objects. \par
    In this work, we present MeerKAT UHF ($813~\mathrm{MHz}$) and L-band ($1.3~\mathrm{GHz}$) observations of the cluster Abell 2244 (A2244, $z = 0.095$), in addition to the previously published observations from the LoTSS.
    The cluster hosts a low-luminosity RH first reported in \cite{Botteon2022a}, while \cite{Balboni2024} show that the RH presents a double component profile but do not give a classification for this emission. 
    In this paper we present an in-depth study of the diffuse radio emission present in the cluster and investigate the possibility of the outer component being related to complex diffuse radio sources, such as megaHs and multi-component haloes. \par
    The outline of the paper is the following: in Sect. \ref{introtarget} we introduce the target; in Sect. \ref{procedure} we present the observations, data reduction, and analysis procedure; in Sect. \ref{results} we present the results and finish with the discussion and conclusions in Sects. \ref{discussion} and \ref{conclusions}. \par
    We assumed a $\Lambda$ cold dark matter cosmological model with $H_0 = 70~\mathrm{km/s/Mpc}$, $\Omega_M = 0.3,$ and $\Omega_\Lambda = 0.7$.
    For the convention of the spectral index, $\alpha$, we used $S_\nu \propto \nu^{-\alpha}$. 
    
\section{Abell 2244}\label{introtarget}
    Abell 2244 is an intermediate-mass cluster \citep[$M_{500} = (4.4 \pm 0.2) \times 10^{14}~\mathrm{M_\odot}$, $z = 0.095$, $r_{500} = 1124 \pm 17~\mathrm{kpc}$;][]{Planck2016} that was observed and studied as part of the LoTSS-\textit{Planck} cluster survey \citep{Botteon2022a}.
    The presence of a two-component radio emission extending up to $800~\mathrm{kpc}$ in the cluster was already reported by \cite{Balboni2024} at $144~\mathrm{MHz}$, showing that a single exponential profile cannot describe the global profile of the diffuse emission. 
    The cluster does not host morphologically complex sources, with the exception of a possible tailed radio galaxy (T1 hereafter) in the north-eastern direction from the cluster centre, which is well visible in MeerKAT UHF- and L-band, and a possible radio plume, visible only in the LOFAR image, located close to the cluster centre and that is not related to any of the nearby sources. 
    \begin{figure}
        \centering
        \includegraphics[width=0.85\linewidth]{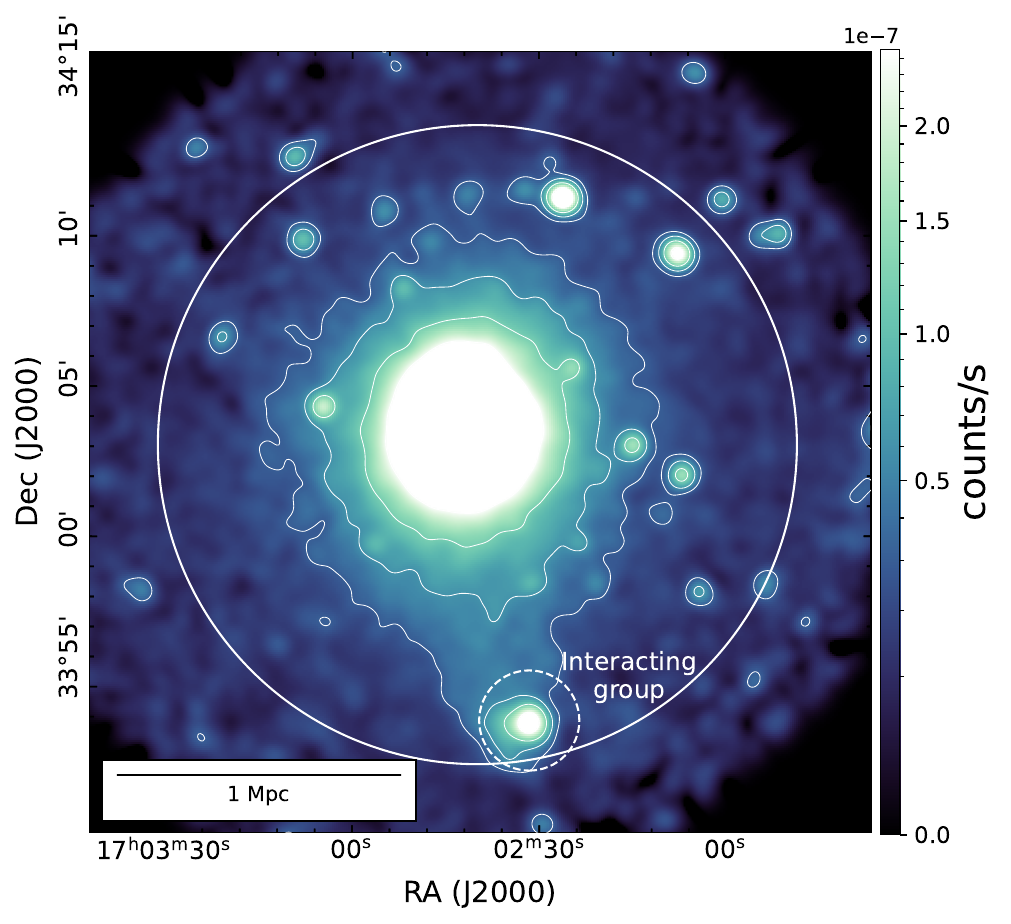}
        \caption{\textit{XMM-Newton} image of A2244 highlighting the presence of the X-ray bridge and the interacting group. The contours are set at $[2, 4, 8] \times \sigma_{rms}$, where $\sigma_{rms} = 1.5 \cdot 10^{-8} \ \mathrm{counts/s}$. The white circle is $r_{500}$.}
        \label{fig:XMM_image}
        \vspace{-3mm}
    \end{figure}
    These sources and their location are highlighted in the LOFAR image in Fig. \ref{fig:sources_imgs} (left panel). \par
    Following the definition in \cite{Cassano2010b}, the cluster has been classified as a mixed morphology cluster by \cite{Campitiello2022}, since, despite having a high concentration and small centroid shift, the ICM distribution shows asymmetries, as shown in Fig. \ref{fig:XMM_image}.
    Using Chandra observations, \cite{Donahue2005} measured the central cooling time of the cluster to be shorter than the Hubble time, classifying it as a cool-core cluster (CC).
    Despite the central cooling time, the central entropy ($K_0 = 48 \pm 5~\mathrm{keV / cm^2}$) and the temperature profiles observed are not consistent with a purely CC, but are intermediate between a relaxed and a disturbed cluster, as the temperature profile is almost isothermal at $r<500~\mathrm{kpc}$ and the entropy profile is flatter than the ones from relaxed clusters.
    These results were recently confirmed by \cite{Andreon2023} using deep Swift-XRT observations.
    X-ray dynamical parameters show the presence of dynamical disturbance in the cluster, for which the most likely candidate is a group ($M_{500} = 2-5 \times 10^{13} \ \mathrm{M_\odot}$) in the southern part of the cluster, which is leaving a trail of gas in the same direction \citep{Balboni2024}.
    The location of the group is highlighted in the bottom left panel of Fig. \ref{fig:sources_imgs} and in Fig. \ref{fig:XMM_image}.
    Considering the combination of dynamical and morphological properties, the cluster can be classified as an almost relaxed cluster with a warm core.
    This is further confirmed when considering the central entropy of the cluster in the mass-entropy diagram showed by \cite{Giacintucci2017} in Fig. 8, as A2244 ends up in the intermediate region between CC and non-CC clusters.
    \begin{figure*}[!htb]
        \centering
        \includegraphics[width=0.32\textwidth]{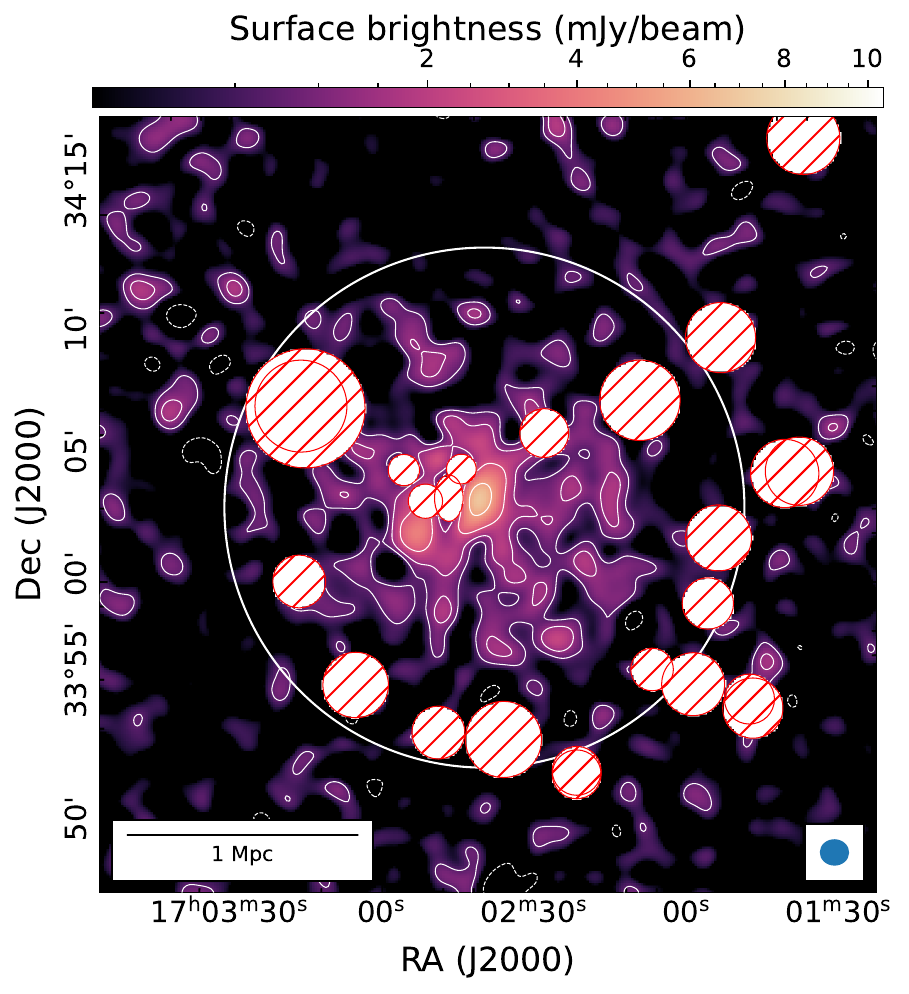}
        \includegraphics[width=0.32\textwidth]{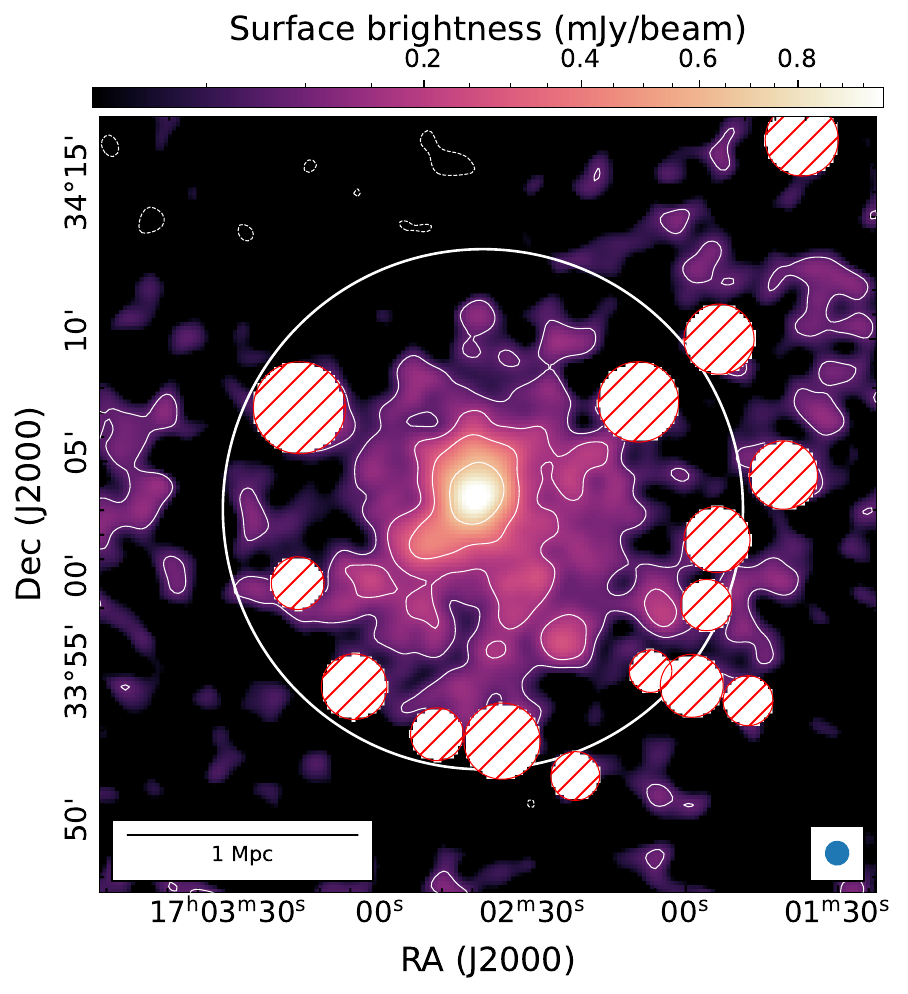}
        \includegraphics[width=0.32\textwidth]{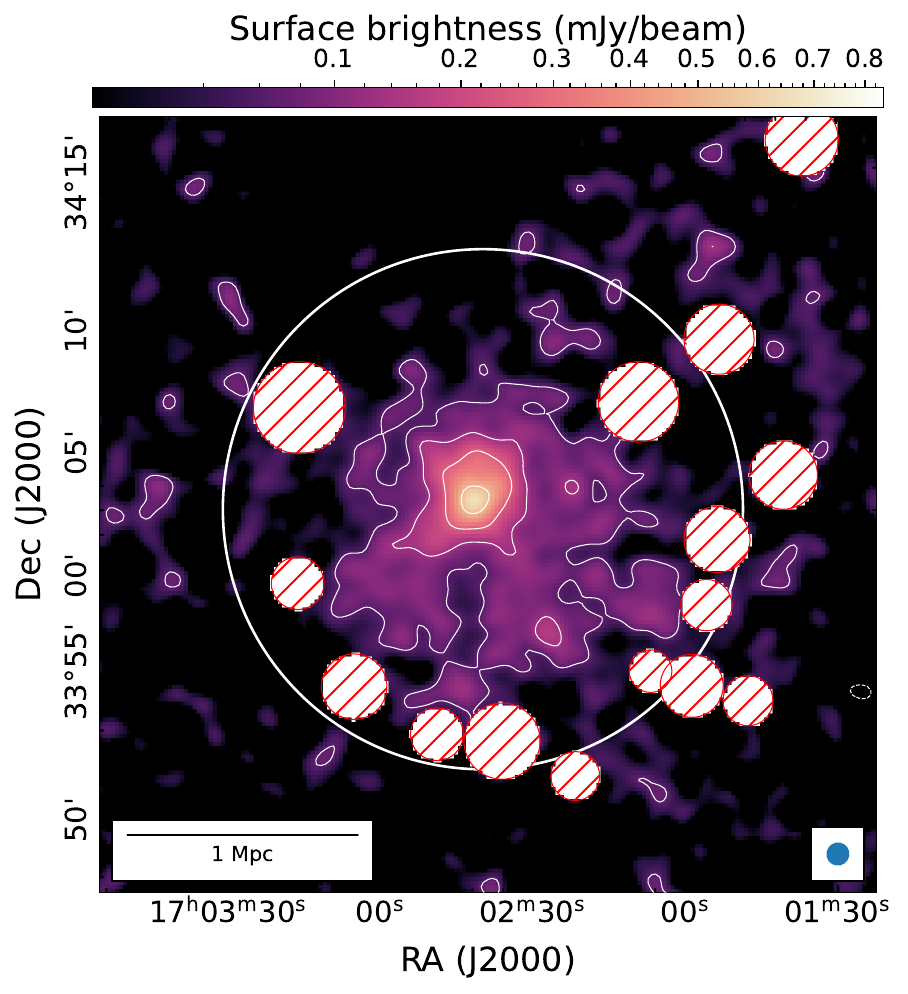}
        \caption{Low-resolution images of A2244 in LOFAR HBA (\textit{left}), MeerKAT UHF (\textit{middle}), and MeerKAT L (\textit{right}) bands. The beam sizes are $72.1\arcsec \times 64.8\arcsec$, $61.2\arcsec \times 60.2\arcsec$ , and $57.5\arcsec \times 55.4\arcsec$ , respectively. The contours are set at levels of $[-3, 2,4,8,16] \cdot \sigma_{rms}$, where $\sigma_{rms} = 340 \ \mathrm{\mu Jy/beam}$, $\sigma_{rms} = 40 \ \mathrm{\mu Jy/beam,}$, and $\sigma_{rms} = 32 \ \mathrm{\mu Jy/beam}$ for HBA, UHF, and L-band. The beam size is shown in the bottom right corner of the images. We show the masks used to remove the contribution from subtraction artefacts. The white circle is $r_{500}$.}
        \label{fig:A2244_imgs}
        \vspace{-3mm}
    \end{figure*}
    \begin{figure*}[!htb]
        \centering
        \includegraphics[width=0.33\textwidth]{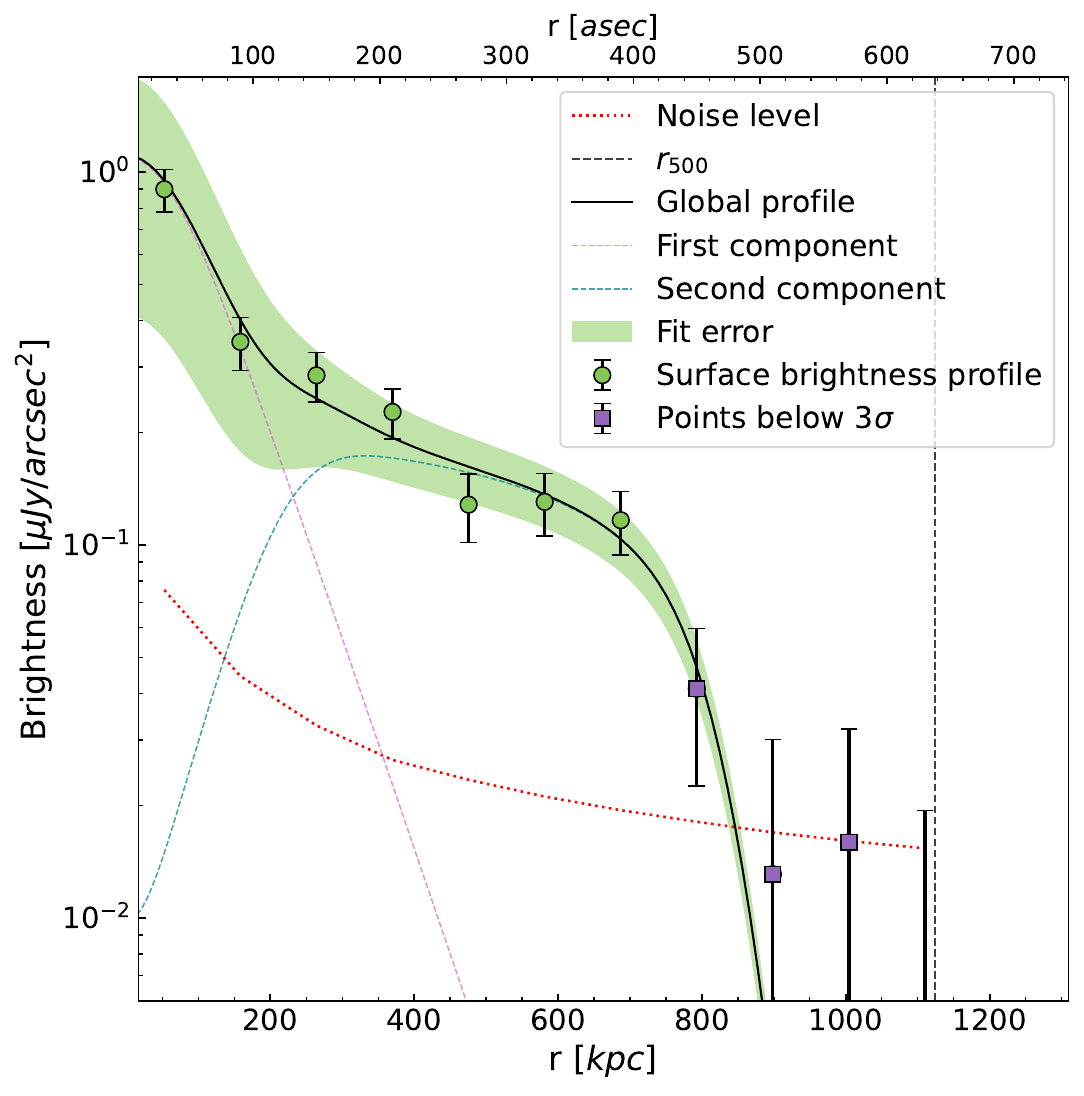}
        \includegraphics[width=0.33\textwidth]{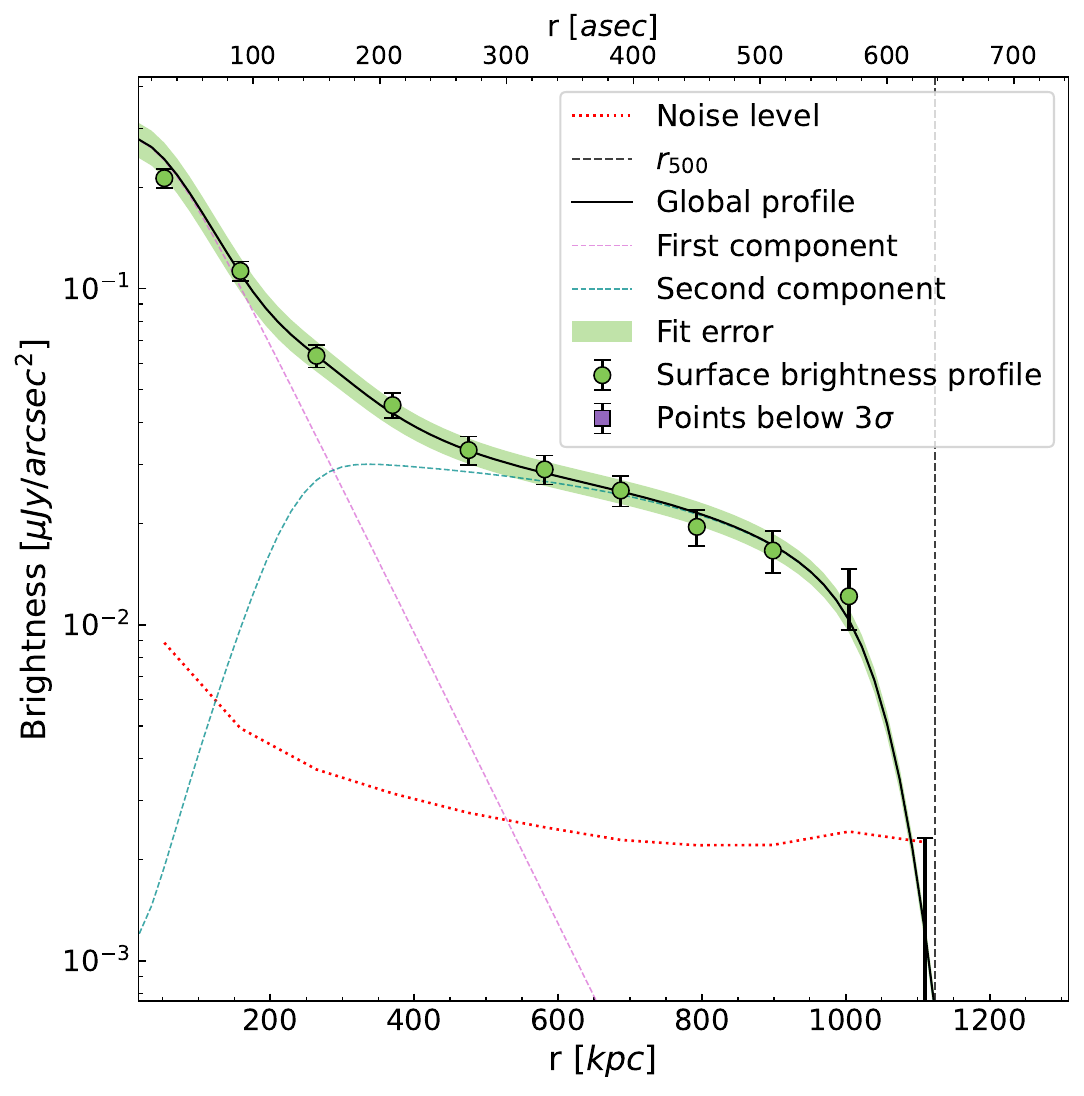}
        \includegraphics[width=0.33\textwidth]{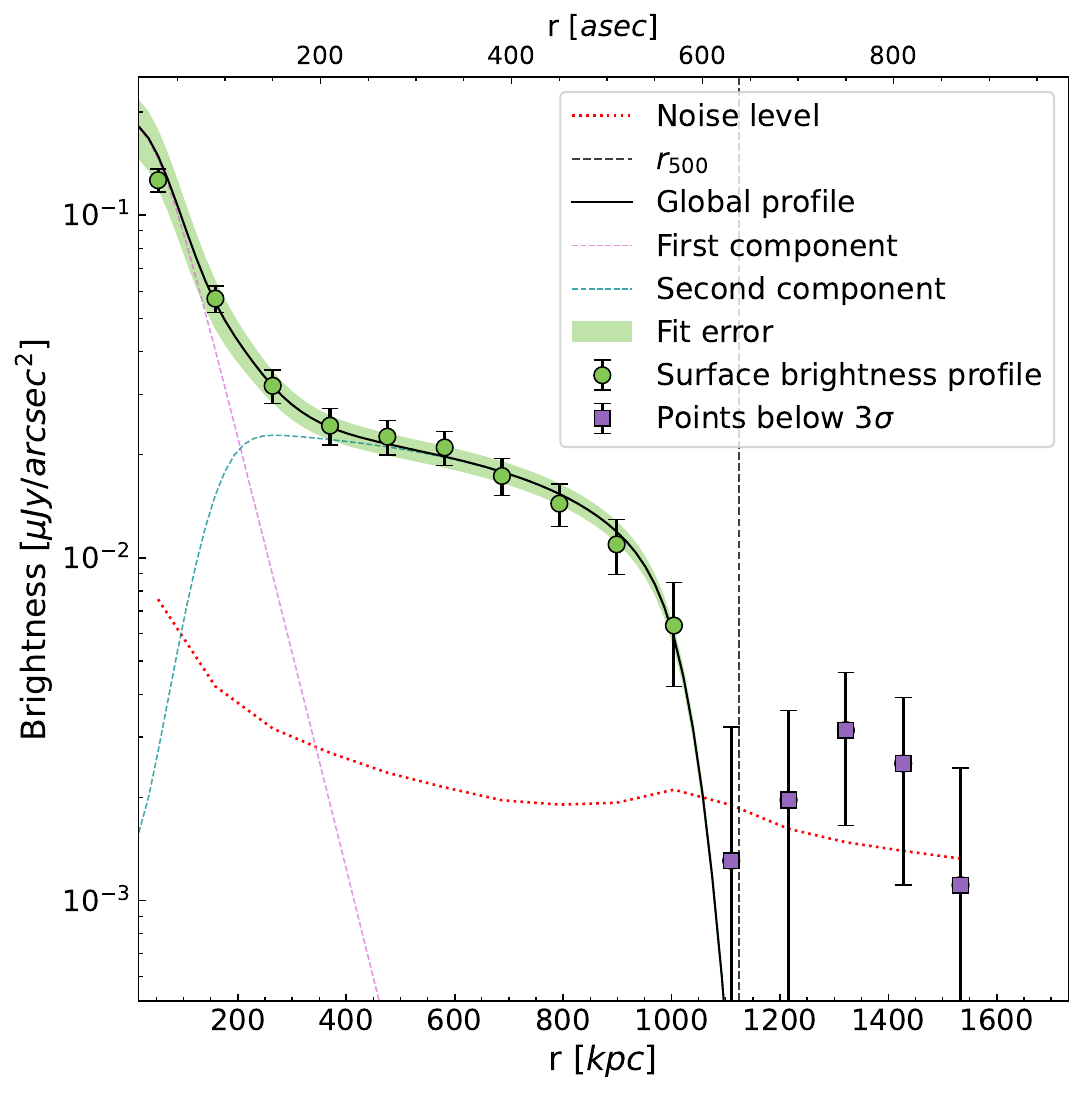}
        \caption{Surface brightness profiles at $144~\mathrm{MHz}$ (\textit{left}), $813~\mathrm{MHz}$ (\textit{middle}), and $1279~\mathrm{MHz}$ (\textit{right}) of the diffuse emission. The solid black line is the fitted profile. The dotted red line is the $1\sigma$ detection limit in each annulus. The vertical dashed line is $r_{500}$.}
        \label{fig:A2244_profs}
        \vspace{-3mm}
    \end{figure*}

\section{Data and methods}\label{procedure}
    In this work, we re-analysed the data of A2244 of the LoTSS Data Release 2 \citep{Shimwell2022, Botteon2022a} and analysed MeerKAT UHF and L-band data.
    The cluster falls within two LoTSS pointings (frequency range $120-168~\mathrm{MHz}$) of eight hours each, that were presented in \cite{Botteon2022a} and analysed in depth in \cite{Balboni2024}.
    The MeerKAT UHF band ($544-1088~\mathrm{MHz}$) observation consists of two pointings of 2.5 hours each on the target (project code: SCI-20220822-MB-04, PI: M. Balboni), while for MeerKAT L-band ($856-1712~\mathrm{MHz}$) we have a single 4.5 hours observation (project code: SCI-20220822-RV-01, PI: R. J. van Weeren). \par
    To carry out the point-to-point radio-X-ray analysis we used data from the Cluster HEritage project with \textit{XMM-Newton} – Mass Assembly and Thermodynamics at the Endpoint of structure formation \citep[CHEX-MATE;][]{CHEX-MATE2021}, totalling $22.7~\mathrm{ks}$ of observation time, which allows a uniform mapping of the cluster X-ray emission within $r_{500}$.
    These data were already presented and discussed in \cite{Balboni2024}.

\subsection{Calibration}
In the following subsections we describe the data calibration methods for the radio observations, divided between LOFAR and MeerKAT, and the data reduction procedure followed for \textit{XMM-Newton} data.
The radio data are compressed using the \textsc{Dysco} compression \citep{Offringa2016}
    \begin{figure*}[!htb]
        \centering
        \includegraphics[width=0.43\textwidth]{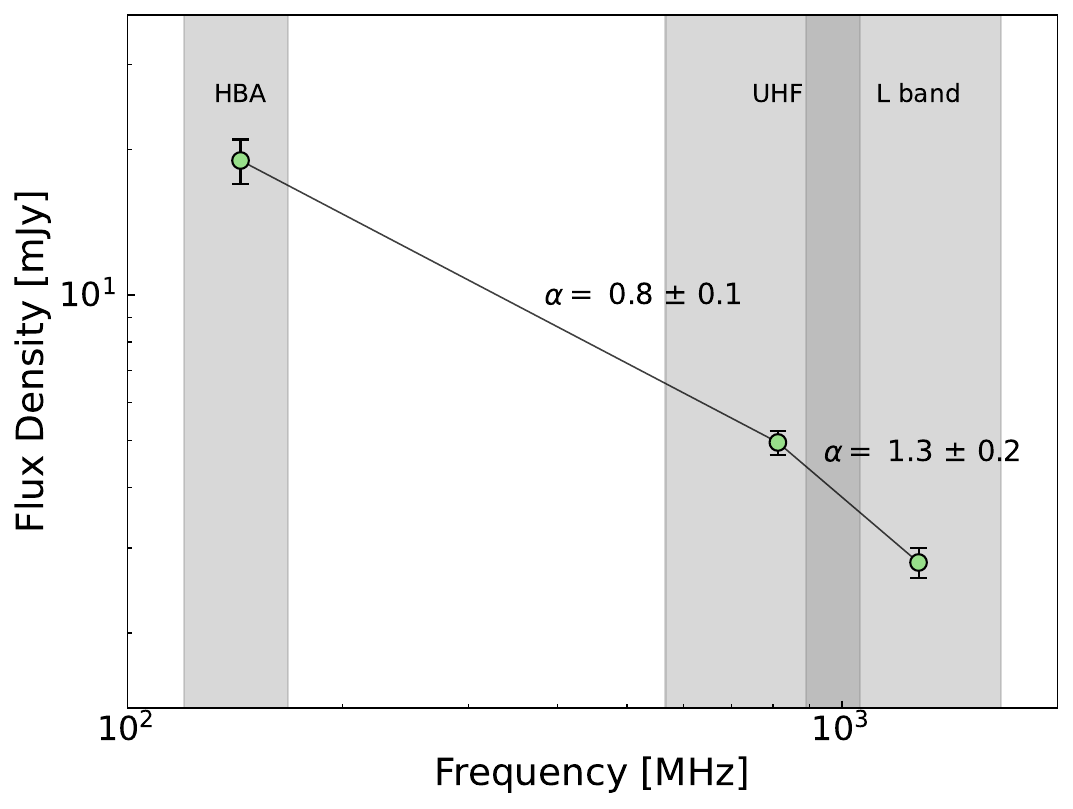}
        \includegraphics[width=0.43\textwidth]{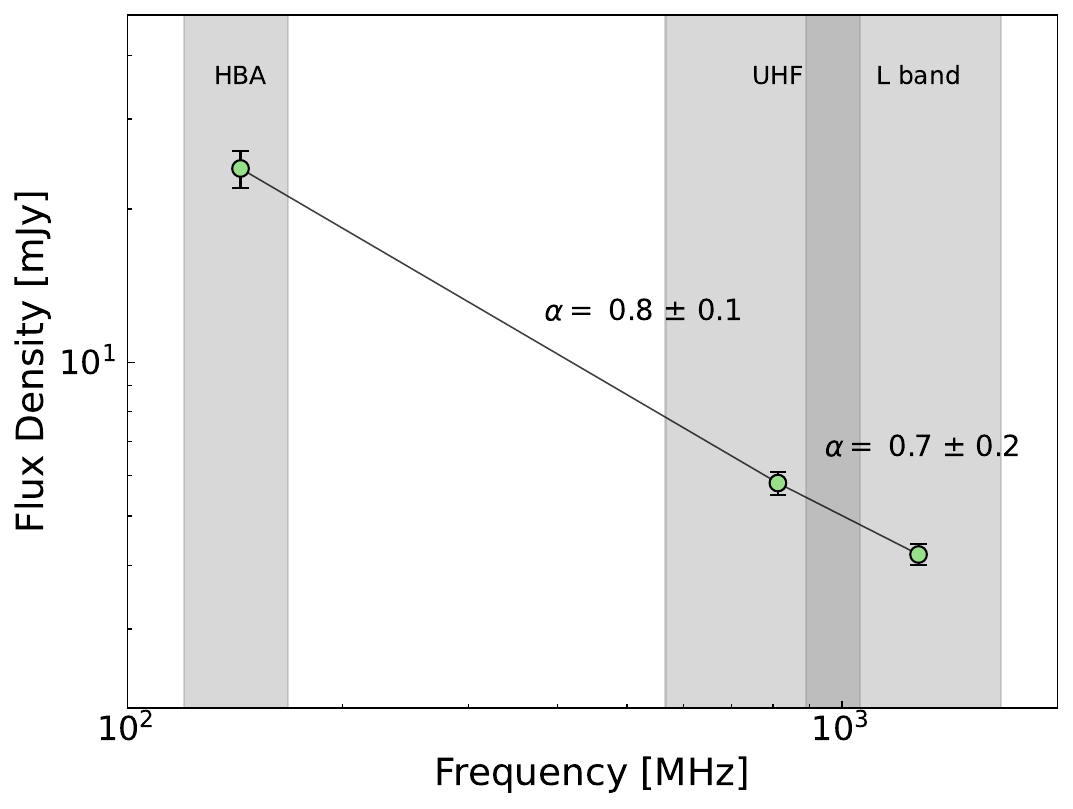}
        \caption{Integrated spectrum of the RH (\textit{left}) and of the outer component (\textit{right}) in A2244 between $144~\mathrm{MHz}$ and $1279~\mathrm{MHz}$. The spectral index for the RH was computed in an area with radii between zero and $\sim 300~\mathrm{kpc}$, while for the outer component the area has an inner radius of $\sim 450~\mathrm{kpc}$ and outer radius of $\sim 800~\mathrm{kpc}$. The regions are shown in Fig. \ref{appendix_fig:flux_regions}.}
        \label{fig:radiohalo_spix}
        \vspace{-3mm}
    \end{figure*} 
\subsubsection{LOFAR data calibration}
    The LOFAR data are calibrated and reduced using the LOFAR Surveys Key Science Project data reduction pipeline \citep{Tasse2021}, which corrects for both direction independent and direction-dependent effects \citep{Smirnov2015, vanWeeren2016, Williams2016, Tasse2018, deGasperin2019}.
    To further improve the data quality and the calibration solutions on the target, we extracted and self-calibrated the visibilities using \texttt{facetselfcal}\footnote{\url{https://github.com/rvweeren/lofar_facet_selfcal/}}, described in detail in \cite{vanWeeren2021}.
    The scheme consists in a first subtraction of all the sources outside of a square region of area $\lesssim 1 \ \deg^2$ containing the target, then, the direction-dependent solutions derived in the calibration step are applied and the phase centre is shifted to the centre of the region, after the extraction, various rounds of self-calibration are applied on the target.
    The software takes advantage of \texttt{DP3} \citep{vanDiepen2018, Dijkema2023} to calibrate the data and \textsc{WSClean} \citep{Offringa2014, Offringa2017} for imaging.
    In our case, we extracted the target within a $0.5 \deg \times 0.5 \deg$ box and we ran ten cycles of self-calibration on the target, of which four with phase solutions and the last six with phase and amplitude calibration solutions.
    For more in depth explanation of the calibration procedure we refer to \cite{Botteon2022a} and \cite{Shimwell2022}.

\subsubsection{MeerKAT data calibration}
    The calibration of MeerKAT observations is divided in three main steps: 1. application of the Science Data Processor (SDP)\footnote{\url{https://skaafrica.atlassian.net/wiki/spaces/ESDKB/pages/338723406/SDP+pipelines+overview}} Calibration Pipeline solutions; 2. self-calibration of the data on the full field of view; 3. extraction and self-calibration on the target. 
    For the first step, we downloaded the data from the MeerKAT archive using the `default calibration' option, which corrects the visibilities for bandpass, delay and gain calibration \citep{Hugo2021} and then converts the visibilities into \textsc{CASA} Measurement Set format \citep{McMullin2007, Casa2022} using the \texttt{mvftoms.py} script of the \texttt{katdal}\footnote{\url{https://github.com/ska-sa/katdal}} package.
    Afterwards, we compressed the data using \textsc{Dysco} and averaged them in time and frequency by a factor of two.
    For the last two steps we used \texttt{faceselfcal}, where in the first part we self-calibrated the data on the full field of view, after an initial round of flagging on the Stokes V visibilities \citep{Botteon2024} using \textsc{AOFlagger} \citep{Offringa2010, Offringa2012} to remove radio frequency interference, and then extracted a region around the target for the second round of self-calibration. \par
    For MeerKAT UHF observations we performed four self-calibration cycles, of which two are phase-only and two phase and amplitude corrections, on the full field of view using \texttt{faceselfcal}.
    Then, we extracted the dataset in a box of $0.5 \deg \times 0.5 \deg$ centred on the target and performed four more rounds of self-calibration with the same steps used before the extraction.
    The procedure is explained more in depth in \cite{Botteon2024} and \cite{Balboni2025}.
    For MeerKAT L-band, we carried out three iterations of phase-only corrections, then we extracted the data in the same box used for the other observations, and performed three more rounds of self-calibration correcting for both phase and amplitude.

\subsubsection{\textit{XMM-Newton} data reduction}
    The data reduction of the \textit{XMM-Newton} X-ray data form the CHEX-MATE sample has been discussed in depth in \cite{Bartalucci2023}, we go over the main steps in this section.
    The target was observed using the European Photon Imaging Camera \citep[EPIC;][]{Turner2001} and the data were analysed using the Extended-Science Analysis System \citep[E-SAS;][]{Snowden2008}.
    To produce the scientific images, the photon-count image was extracted in the $0.7-1.2~\mathrm{keV}$ band for each camera and the exposure map was produced with the tool \texttt{eexpmap}.
    The X-ray background is composed by the local Galactic emission, an instrumental component, and the cosmic X-ray background \citep{Kuntz2000}.
    To maximise the statistics the maps from different cameras were merged together \citep{Bartalucci2023}.
    As a last step, the background map was subtracted from the source map and was then corrected using the exposure map.
    
\subsection{Imaging and source subtraction}\label{source_sub}
    Our imaging and source subtraction approach is based on the one followed by \cite{Cuciti2022}.
    We produced a first image with an inner uv-cut of $80\lambda$ at high resolutions ($6\arcsec$ for LOFAR HBA, $8\arcsec$ for MeerKAT L-band, and $10\arcsec$ for MeerKAT UHF) using the \cite{Briggs1995} scheme with \texttt{robust = -0.5}, and investigated the images by eye to search for the presence of complex discrete radio sources that could cause issues in the subtracted images if not properly masked; these images are shown in Fig. \ref{fig:sources_imgs}.
    After this, we produced high-resolution images with an inner uv-cut of $1.5 k\lambda$ (corresponding to $250~\mathrm{kpc}$ at the cluster's redshift) and subtracted the derived model from the visibilities.
    This allows us to subtract the sources without affecting the diffuse radio emission.
    To check the quality of the subtraction, we produced intermediate-resolution images (with a Gaussian taper of $25\arcsec$) with an inner uv-cut of $80\lambda$ to highlight subtraction residuals, and found only part of the tail of T1 to still be present. These images are shown in Fig. \ref{fig:r500_20asec}. 
    We also observe diffuse emission close to the group centre in the south, which we associate with the presence of subtraction residuals of the radio galaxies in the region.
    After confirming the good quality of the source subtraction, we produced low-resolution images to highlight the diffuse radio emission.
    To retrieve most of the diffuse emission, we used again an inner uv-cut of $80\lambda$ and applied a Gaussian taper of $60\arcsec$.
    We produced the images using \textsc{WSClean}, taking advantage of the \texttt{multi-frequency} synthesis and \texttt{multi-scale} cleaning \citep{Offringa2017}.
    
\subsection{Surface brightness profiles}
    Azimuthally averaged radial profiles allow us to search for double component diffuse radio sources in galaxy clusters.
    As the diffuse emission in A2244 is asymmetric, and mostly found in the southern direction of the cluster, creating circular concentric annuli to estimate the surface brightness profile would introduce biases in the resulting profile.
    To avoid this issue, we produced the surface brightness profiles by taking concentric half-annuli, starting at an angle of $180\degree$ and ending at $360\degree$ in an anti-clockwise direction, hence covering the southern region of the cluster, of width equal to the major axis of the beam centred on the brightest pixel of the RH in the low-resolution source-subtracted images.
    To avoid contamination from image artefacts, we masked all the subtraction residuals close to the diffuse emission from the cluster.
    To fit the profile we used a $2$D exponential model for the RH \citep{Murgia2009} and a projected thick shell for the outer component.
    The complete expression is
    \begin{equation}
    I(r) =
    \begin{cases}
        I_0 e^{-r/r_e} + I_1 \left( \frac{\sqrt{R_{out}^2 - r^2}}{R_{out}} - \frac{\sqrt{R_{in}^2 - r^2}}{R_{in}} \right) & \phantom{R_{min} <{}} r < R_{in} \\[2ex]
        I_0 e^{-r/r_e} + I_1 \frac{\sqrt{R_{out}^2 - r^2}}{R_{out}} & R_{in} < r < R_{out},
    \end{cases}
    \label{shell}
    \end{equation}
    where $I_0$ is the central brightness of the RH and $r_e$ is its effective radius, $I_1$ is the central brightness of the outer component, and $R_{in}$ and $R_{out}$ are its inner and outer radius.
    The units of the central brightnesses are in $\mathrm{Jy/arcsec^2}$ and the ones of the radii are in $\mathrm{arcsec}$. \par
    We compared this model with a double exponential one, described as    \begin{equation}
        I(r) = I_1 e^{-r/r_{e1}} + I_2 e^{-r/r_{e2}} \ \ \mathrm{[Jy/arcsec^2]}.
    \end{equation}For the fitting procedure, we produced a $2$D model with the same pixel scale and image size of the images and convolved it with a Gaussian kernel having full width at half maximum equal to the beam size.
    This allows us to take into account the resolution of the image in the model, as explained in \cite{Murgia2009}.
    To find the best-fit results we used a non-linear least squares method, where we compared the surface brightness of the emission and of the model in the same annuli.
    All the annuli with a surface brightness greater than zero have been used to derive the best-fit parameters; the annuli with values closer to the noise level have larger errors and affect the fit proportionally less.
    
\subsection{Flux density and spectral index measurements}\label{measurements}
    We measured the flux density from the images, drawing a region in the location where the RH is dominant over the second component, avoiding the image artefacts, which have been masked, as shown in Fig. \ref{fig:A2244_imgs}.
    We also followed the same procedure  for the outer component, selecting the region where it is dominant over the RH emission.
    The regions are showed in Fig. \ref{appendix_fig:flux_regions} in yellow, with the masks applied to avoid contamination from subtraction residuals. The region used to measure the flux density of the outer component extends from $\sim 450~\mathrm{kpc}$ to $\sim 800~\mathrm{kpc}$, which is where the outer component is detected in all the images.
    The shape of the central region is chosen to carefully avoid the central artefact present in the LOFAR image.
    These flux density measurements are used exclusively to compute the spectral indices for the inner and outer components of the diffuse radio emission. \par
    We derived the spectral indices using the expression
    \begin{equation}
        \alpha = - \ln(S_2/S_1)/\ln(\nu_2/\nu_1),
    \end{equation}
    and the associated error is
    \begin{equation}
        \Delta \alpha = \left| \dfrac{1}{\ln(\nu_2/\nu_1)} \right| \sqrt{\left(\dfrac{\Delta S_1}{S_1}\right)^2 + \left(\dfrac{\Delta S_2}{S_2}\right)^2},
    \end{equation}
    where $\Delta S$ is
    \begin{equation}
        \Delta S = \sqrt{(\sigma_{rms} \cdot \sqrt{N_{beam}})^2 + (S \cdot \delta_{cal})^2},
    \end{equation}
    where $\sigma_{rms}$ is the rms of the image, $N_{beam}$ is the number of beams in the area of interest, and $\delta_{cal}$ is the calibration error.
    For LOFAR HBA we assume $\delta_{cal} = 10\%$ \citep{Shimwell2022, Botteon2022a}, while for MeerKAT we assume $\delta_{cal} = 5\%$ \citep{Perley2013}. \par
    To derive the spectral index of the RH and of the outer component between $144~\mathrm{MHz}$ and $1279~\mathrm{MHz}$, we fitted the spectrum using a power-law relation ($S \propto \nu^{-\alpha}$), with the flux density measurements obtained from the regions shown in Fig. \ref{appendix_fig:flux_regions}, we refer to this one as the integrated spectral index.
    To compute the radio power of the RH and compare it with the results from \cite{Cuciti2023}, we used the analytical flux density, computed using the expression \citep{Murgia2009}
    \begin{equation}\label{exp}
        S_\nu = 2 \pi f I_{\nu,0} r_e^2 \ \ \mathrm{[Jy]},
    \end{equation}
    where $I_{\nu,0}$ is in $\mathrm{Jy/arcsec^2}$, $r_e$ is in $\mathrm{arcsec}$, and $f$ is the fraction of flux density within the integration radius to the total one. 
    As RHs do not have an infinite extension, the computation of their flux density is truncated at an arbitrary radius, which typically is $3r_e$, which corresponds to a value of $f = 0.8$.
    Since the MeerKAT UHF image is the one with the best combination of noise levels and sensitivity to low brightness diffuse emission, as it can be seen from Fig. \ref{fig:A2244_imgs}, we computed the analytical flux density using the best-fit parameters obtained from this image, and then rescaled the flux density to $150~\mathrm{MHz}$ using the measured spectral index.
    From this value, we computed the luminosity using the expression
    \begin{equation}
        P_\nu = 4 \pi S_\nu d_L^2 (1 + z)^{\alpha - 1} \ \mathrm{W/Hz},
    \end{equation}
    where $S_\nu$ is the flux density of the source, $d_L$ its luminosity distance, $\alpha$ the spectral index, and $z$ the redshift of the source.
    We did not use an analytical formula to compute the flux density of the outer component from the fitted profile, as it would assume a symmetric emission and lead to an over-estimation of the flux density.
    
\subsection{Point-to-point correlation}
    Following what has been done by \cite{Govoni2001}, we extracted the average radio and X-ray surface brightnesses from the respective images by creating a grid covering the diffuse emission.
    To separate the surface brightness of the RH from the one of the outer component, we followed the procedure twice. 
    In the first run we selected a box covering only the RH, while in the second one the box covered the area within $r_{500}$ and we masked the RH region.
    In both cases we masked residual emission in the radio image and the positions of the X-ray sources not associated with the ICM diffuse emission that could interfere with the analysis, removing all the cells that were contaminated by the presence of these sources.
    The size of the cells is $70\arcsec \times 70\arcsec$.
    We only considered the boxes where the radio surface brightness was higher than $2\sigma_{rms}$ of the radio image, following \cite{Botteon2020} and \cite{Balboni2024}, and are showed in Fig. \ref{appendix_fig:grids}.
    Before estimating the X-ray surface brightness from each box, we subtracted the background from the source image and corrected it using the exposure map. 
    We only carried out this analysis on the MeerKAT images, as the same analysis on the LOFAR HBA image was carried out by \cite{Balboni2024}.
    We show the resulting point-to-point correlation for MeerKAT UHF band in Fig. \ref{fig:ptp-corr} and for MeerKAT L band in Fig. \ref{fig:ptp-lband}.
    We plotted the X-ray surface brightness ($I_X$) against the radio surface brightness ($I_R$), separating the points from the RH region and the ones from the outer component.
    Then, we fitted the RH relation with a linear relation of the form
    \begin{equation}\label{ptprelation}
        \log I_R = B \log I_X + A,
    \end{equation}
    where $B$ is the slope of the relation.
    The slope allows us to discriminate the type of central diffuse emission, as RHs have $B \lesssim 1$ (sub-linear) and mHs have $B \gtrsim 1$ (super-linear). 
    To fit the relation we used the Bivariate Correlated Errors and intrinsic Scatter (BCES) linear regression algorithm \citep{Akritas1996} using the orthogonal method, taking advantage of the \textsc{BCES}\footnote{\url{https://github.com/rsnemmen/BCES}} Python package \citep{Nemmen2012}.
    
\section{Results}\label{results}
In this section we present the results of the study, divided into four subsections. In the first one we report the flux densities, radio powers, and spectral indices at the different frequencies, in the second one we focus on the spectral index map and spectral index radial profile, in the fourth section we focus on the secondary tests related to the surface brightness profiles, and in the last section we overview the results from the point-to-point radio X-ray relation.
\subsection{Diffuse radio emission}\label{results_diffuse}
    {\renewcommand{\arraystretch}{1.15}
    \begin{table*}[!htb]
        \caption{Results of the fit of the diffuse emission. }
        \centering
        \begin{tabular}{ccccccc}
            \hline \hline
            Frequency & $I_0$ & $r_e$ & $I_1$ & $R_{in}$ & $R_{out}$ & $\tilde{\chi}^2$ \\
            $\mathrm{[MHz]}$ & $\mathrm{[\mu Jy/arcsec^2]}$ & [kpc] & $\mathrm{[\mu Jy/arcsec^2]}$ & [kpc] & [kpc] &  \\
            \hline
            144 & $2.4 \pm 1.5$ & $76 \pm 54$ & $0.20 \pm 0.04$ & $238 \pm 274$ & $826 \pm 38$ & $1.05$ \\ 
            \hline
            813 & $0.48 \pm 0.06$ & $100 \pm 14$ & $0.032 \pm 0.003$ & $244 \pm 110$ & $1073 \pm 29$ & $0.38$ \\
            \hline
            1279 & $0.37 \pm 0.07$ & $68 \pm 14$ & $0.024 \pm 0.002$ & $177 \pm 91$ & $1046 \pm 32$ & $1.1$ \\
            \hline
        \end{tabular}
        \label{tab:fit_table}
        \tablefoot{From left to right we have: frequency of the observation, central brightness of the RH $I_0$, effective radius of the RH $r_e$, central brightness of the outer component $I_1$, inner radius of the outer component $R_{in}$, outer radius of the outer component $R_{out}$, and reduced $\chi^2$ ($\tilde{\chi}^2$) of the model. The $\tilde{\chi}^2$ of the fit at $813~\mathrm{MHz}$ is smaller than the one at $1279~\mathrm{MHz}$ despite the radial profiles being similar, as the fit is carried out over all the points of the profiles, also considering the ones that are below the $3\sigma$ detection.}
        \vspace{-3mm}
    \end{table*}
    }
    In Fig. \ref{fig:A2244_imgs} we show the source-subtracted low-resolution images of A2244, while the respective surface brightness profiles are shown in Fig. \ref{fig:A2244_profs}.
    From the images we can observe the presence of the two-component diffuse radio emission, which is highlighted in the surface brightness radial profiles.
    The radial profiles show that most of the detected emission, in terms of spatial coverage, comes from the outer component, which reaches almost $r_{500}$ in the profiles at $813~\mathrm{MHz}$ and $1279~\mathrm{MHz}$.
    The outer component at $144~\mathrm{MHz}$ is less extended with respect to the high-frequency images due to the presence of image artefacts limiting the dynamic range.
    In Appendix \ref{appendix_20asec} we show the intermediate-resolution images, created with a Gaussian taper of $25\arcsec$.
    From these images, we clearly see that the residuals in the MeerKAT images are located near T1, and not close to the centre of the cluster.
    Since the outer component has similar morphological properties and the same extension between UHF and L-band, we can reach the conclusion that we are detecting the same emission in the two bands, and it is not produced by subtraction residuals in the two different images.
    In the LOFAR image, the RH has a smaller extension with respect to the results at higher frequencies, this is due to the presence of a strong patch of negative emission close to the centre of the RH, which affects the morphology and extension of the emission.
    In Tab. \ref{tab:fit_table} we reported the results from the best-fit. \par
    Since we do not recover emission beyond $\sim 800~\mathrm{kpc}$ in the LOFAR data, we decided to only measure the spectral index within this radius. However, there is also a possibility that some fraction of the flux is not recovered by LOFAR even within $800~\mathrm{kpc}$. To investigate this, we rescaled the model image obtained at $813~\mathrm{MHz}$ to $144~\mathrm{MHz}$, using the measured spectral index between the two MeerKAT frequencies, and injected it into an empty region in the LOFAR image (see e.g. \cite{Srikanth2026}).
    To estimate the flux density loss, we took regions similar to the ones used to estimate the flux densities of the RH and of the outer component (shown in Fig. \ref{appendix_fig:flux_regions}), extending from $\sim 450~\mathrm{kpc}$ to $\sim 800~\mathrm{kpc}$.
    The RH flux density of the injected rescaled model is $S_\nu = 25 \pm 4~\mathrm{mJy}$, while the flux density of the outer component is $S_\nu = 22 \pm 4~\mathrm{mJy}$.
    The recovered fluxes from the image, after the injection, are $S_\nu = 24 \pm 3~\mathrm{mJy}$ and $S_\nu = 21 \pm 2~\mathrm{mJy}$ for the RH and outer component respectively. 
    As the injected and recovered flux density are consistent within the errors, we can assume that there is no flux density loss in the regions of interest for our study.
    Since the injected and recovered flux densities are consistent with the measured flux density from the emission at $ 144~\mathrm{MHz}$, we can assume that these regions are not strongly affected by artefacts, hence the flux densities can be used to determine the spectral index of the emission in the region. As the emission is not recovered beyond $\sim 800~\mathrm{kpc}$, we do not measure the flux densities between $\sim 800~\mathrm{kpc}$ and $\sim 1100~\mathrm{kpc}$. The spectral index beyond $\sim 800~\mathrm{kpc}$ is only measured via the spectral index radial profile, computed between the UHF and L bands. \par
\begin{figure}
        \centering
        \includegraphics[width=0.85\linewidth]{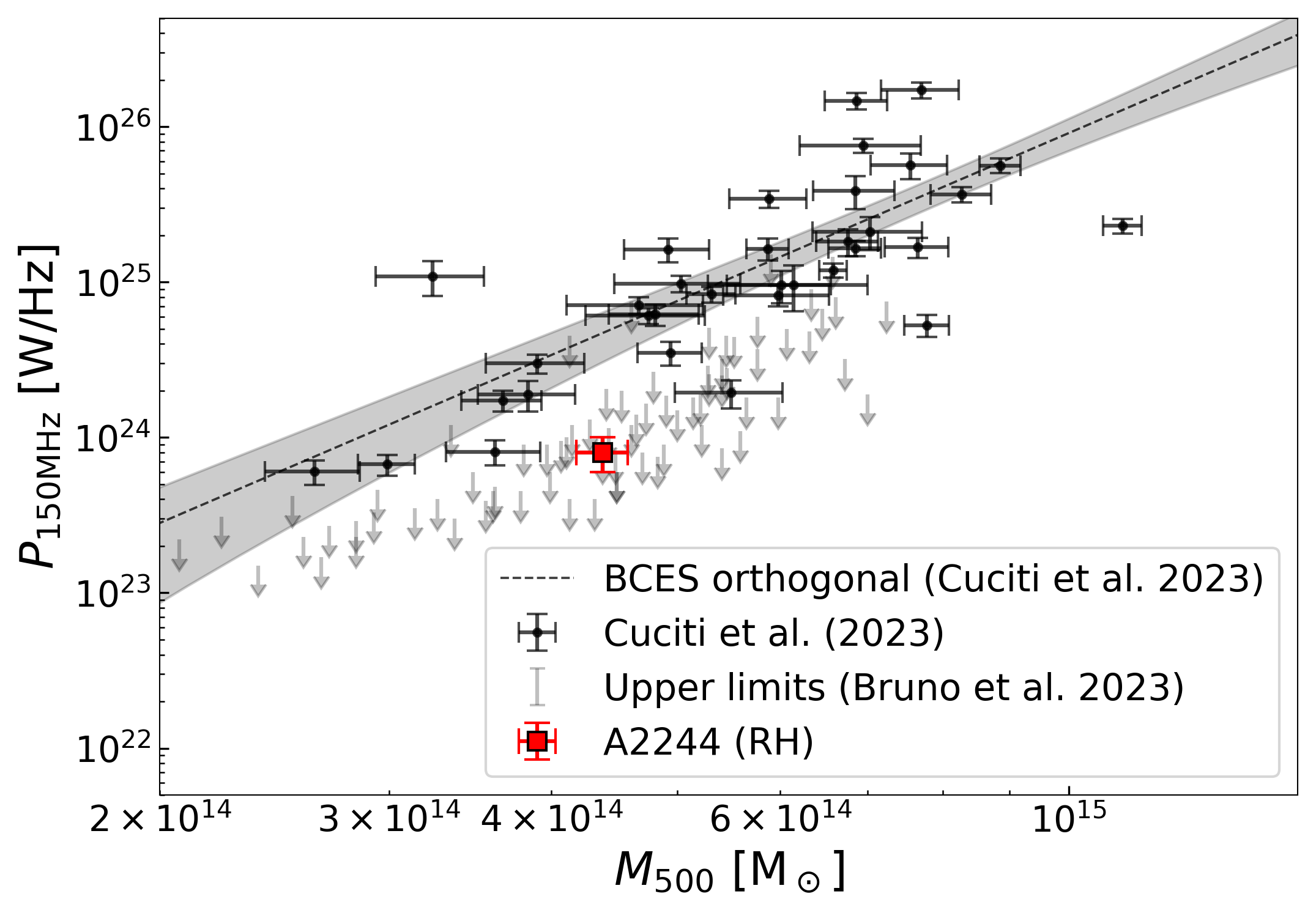}
        \caption{Radio power-mass relation for the \textit{Planck} clusters of the LoTSS-DR2 \textit{Planck} Survey. The black points and the dashed line are the powers and best-fit relation reported by \cite{Cuciti2023}. In red we show the radio power at $150~\mathrm{MHz}$ of A2244 considering only the RH. The grey arrows show the upper limits derived by \cite{Bruno2023}.}
        \label{fig:PM}
        \vspace{-3mm}
    \end{figure}

    The flux densities of the two components of the diffuse emission were computed as explained in Sect. \ref{measurements}.
    For the RH, we obtain $S_{144\mathrm{MHz}} = 19 \pm 2 \ \mathrm{mJy}$, $S_{813\mathrm{MHz}} = 5.0 \pm 0.3 \ \mathrm{mJy}$, and $S_{1279\mathrm{MHz}} = 2.8 \pm 0.2 \ \mathrm{mJy}$, which we used to estimate the spectral indices.
    We first fitted a power-law model to the integrated spectrum, obtaining an integrated spectral index of $\alpha^{1279}_{144} = 0.9 \pm 0.1$. 
    Then we estimated the spectral indices in the ranges $144-813~\mathrm{MHz}$ and $813-1279~\mathrm{MHz}$, that result in $\alpha^{813}_{144} = 0.8 \pm 0.1$ and $\alpha^{1279}_{813} = 1.3 \pm 0.2$.
    The spectrum of the RH is shown in the left panel of Fig. \ref{fig:radiohalo_spix}.
    The analytical flux density of the RH, computed by integrating the best-fit parameters obtained from the image at $813~\mathrm{MHz}$, is $S_{813\mathrm{MHz}} = 7.8 \pm 2.4~\mathrm{mJy}$.
    We rescaled this flux density to $150~\mathrm{MHz}$ using the integrated spectral index, resulting in $S_{150\mathrm{MHz}} = 36 \pm 11~\mathrm{mJy}$. 
    From this flux density we compute the radio power and compare it with the RHs from \cite{Cuciti2023}. We chose to rescale the analytical flux density computed at $813~\mathrm{MHz}$, as this is the most reliable fit obtained, showed by the value of $\tilde{\chi}^2$.
    The corresponding radio power is $P_{150\mathrm{MHz}} = (8.1 \pm 2.5)\cdot 10^{23}~\mathrm{W/Hz}$.
    The RH is under-luminous with respect to the expected value from the radio power-mass relation observed at $150~\mathrm{MHz}$ \citep[Fig. \ref{fig:PM};][]{Cuciti2023}. \par
    To estimate the spectral indices for the outer component we followed the same procedure.
    The measured flux densities, in the region shown in Fig. \ref{appendix_fig:flux_regions}, are $S_{144\mathrm{MHz}} = 24 \pm 2~\mathrm{mJy}$, $S_{813\mathrm{MHz}} = 5.8 \pm 0.3~\mathrm{mJy}$, and $S_{1279\mathrm{MHz}} = 4.2 \pm 0.2~\mathrm{mJy}$.
    The integrated spectral index for the outer component is $\alpha^{1279}_{144} = 0.80 \pm 0.04$, which is consistent with the value found for the RH ($\alpha^{1279}_{144,RH} = 0.9 \pm 0.1$).
    We measure $\alpha^{813}_{144} = 0.8 \pm 0.1$ and $\alpha^{1279}_{813} = 0.7 \pm 0.2$.
    The spectral index is comparable with the one of the RH at lower frequencies, showing a lack of spectral steepening above $813\mathrm{MHz}$.
    Due to the lack of an expression to derive the analytical flux density of the outer component while also taking into account its asymmetric morphology, we decided to not derive its radio power.
    The fact that the outer component has a spectral index comparable to the one of the RH at low frequencies (between $144~\mathrm{MHz}$ and $813~\mathrm{MHz}$), may hint to the possibility that the two sources are originated by the same event, which could be the interaction between A2244 and the nearby group. \par
    It is important to consider that the region in which we measure the flux densities for the RH is small compared to the total size of the RH obtained from the fit.
    This makes the measurement of the integrated spectral index only correct in the region used to measure the flux densities.
    In addition, measuring the flux directly from the images does not allow us to disentangle the two components, meaning that the flux densities can be contaminated in the two used regions.
    To avoid this issue, one should use the analytical flux densities computed using Eq. \ref{exp}, which we cannot do in this case due to an image artefact partially affecting the fit of RH at $144~\mathrm{MHz}$.
    \begin{figure}
        \centering
        \includegraphics[width = 0.8\linewidth]{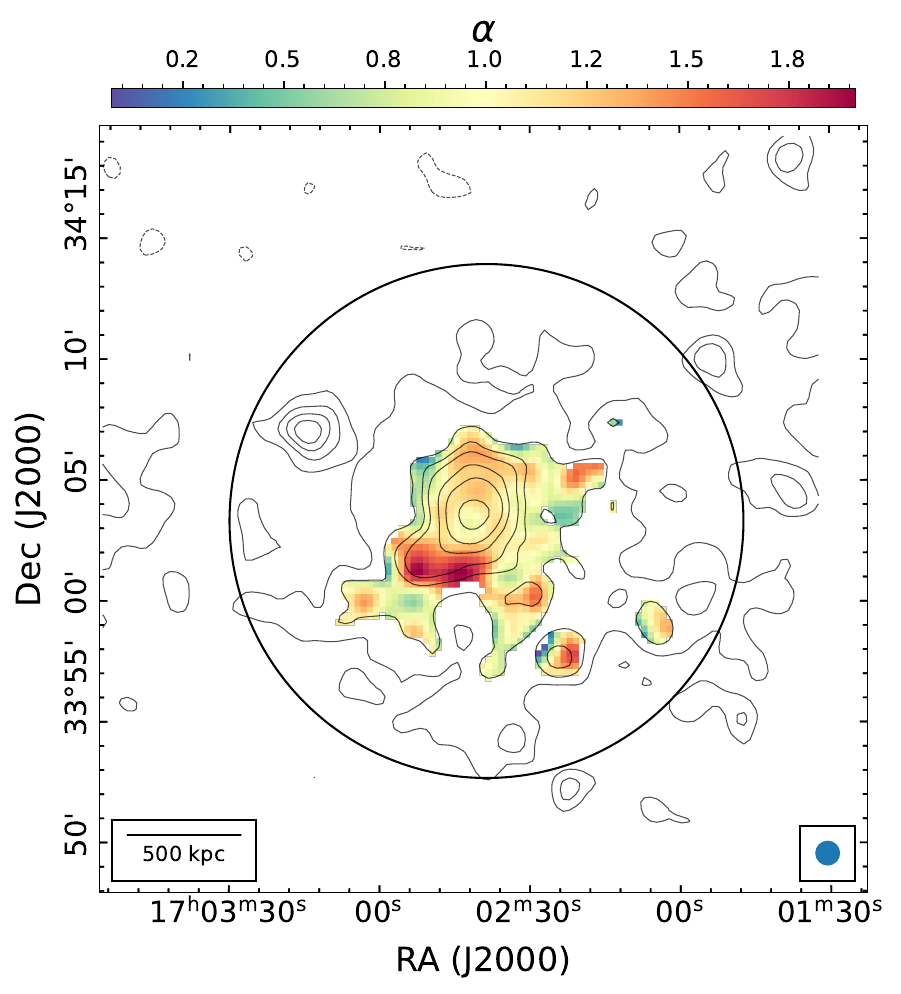}
        \caption{Spectral index map between $813~\mathrm{MHz}$ and $1279~\mathrm{MHz}$. The beam size is $61.8\arcsec \times 61.8\arcsec$. The contours are from the MeerKAT UHF low-resolution source-subtracted image and are set at $[2, 4, 8, 16, 32, 64] \times \sigma_{rms}$, where $\sigma_{rms} = 40 \ \mathrm{\mu Jy/beam}$. The black circle is $r_{500}$.}
        \label{fig:spix_maps}
        \vspace{-3mm}
    \end{figure}
    
\subsection{Spectral index}
    We produced the spectral index map of the diffuse radio emission using only the MeerKAT images, as in the LOFAR HBA image the outer component is not fully detected at a $3\sigma_{rms}$ level.
    To make the spectral index map, we convolved both MeerKAT images to the same resolution ($61.8\arcsec \times 61.8\arcsec$) and set the same inner uv-cut of $80\lambda$.
    The spectral index map is shown in Fig. \ref{fig:spix_maps}, while the related error map is shown in Fig. \ref{fig:spix_errmaps}.
    The outer component is only detected in a small portion of the spectral index map, which prevents us from observing the spectral index distribution over the entire diffuse emission.
    We also notice that these regions have large errors, making the values less reliable than the ones from the RH. \par
\begin{figure}
        \centering
        \includegraphics[width = 0.85\linewidth]{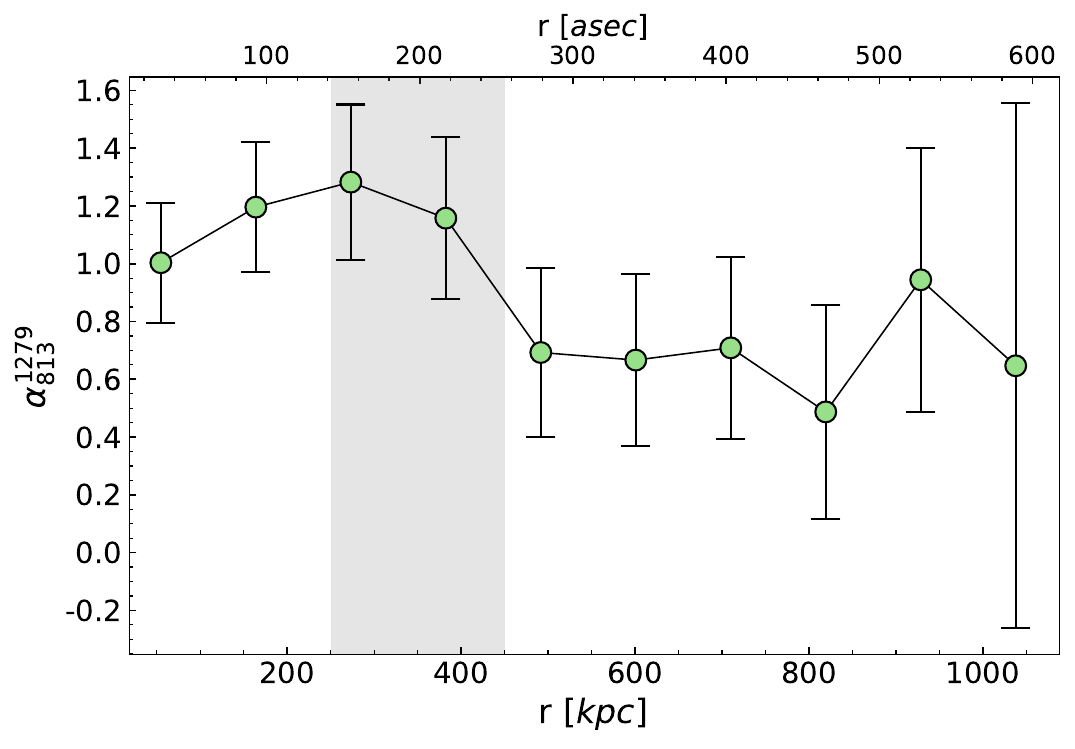}
        \caption{Spectral index radial profile between $813~\mathrm{MHz}$ and $1279~\mathrm{MHz}$. The grey shaded area shows the transition region from the exponential profile to the outer component.}
        \label{fig:profile_spix}
        \vspace{-3mm}
    \end{figure}

    To further investigate the spectral index properties we produced a spectral index profile between UHF and L-band using the same annuli used to produce the surface brightness radial profile and it is shown in Fig. \ref{fig:profile_spix}.
    From the spectral index radial profile, we can see a radial steepening in the RH, while the spectral index of the outer component is flat over a scale of a few hundred kiloparsecs.
    Overall, we see that, due to the large errors, the spectral index of the outer component remains consistent with the one of the RH.
    Such large uncertainties on the spectral index do not allow us to put stringent constraints on the nature and origin of the acceleration mechanisms of the emission. \par

\subsection{Surface brightness profiles}\label{profiles}
    From Fig. \ref{fig:A2244_profs}, we see how the diffuse radio emission in A2244 shows an outer component, which is well fitted by the model reported in Eq. \eqref{shell}.
    We compared our model with a double exponential one.
    The resulting best-fitted profiles are shown in Appendix \ref{appendix_double_exp}.
    As we can see from Fig. \ref{fig:double_exp_prof}, the double exponential model can properly fit the observed profiles, with the exception for the MeerKAT UHF image, in which the outer component displays a cut-off at $r_{500}$.
    To understand which model is better between the shell and the double exponential, we computed the $\tilde{\chi}^2$ for every fit at different frequencies.
    These values are reported in Tab. \ref{tab:fit_table} for the shell model and in Tab. \ref{tab:fit_table_double_exp} for the double exponential.
    We see that, in every fitting run, the shell model performs better than the double exponential model. \par
    To confirm the actual presence and existence of the large-scale diffuse emission around the RH, we followed a similar procedure to the one reported by \cite{Rajpurohit2025} in order to check for contribution from faint point sources or post-subtraction residual emission.
    We produced intermediate $25\arcsec$ resolution source-subtracted images and extracted the surface brightness profiles after masking all the sources present in the source catalogue made from the MeerKAT UHF image, the masked and un-masked images are shown in Fig. \ref{fig:r500_20asec}.
    In Fig. \ref{fig:sb_comparison} we show the profiles obtained from the low-resolution $60\arcsec$ and intermediate-resolution $25\arcsec$ images.
    We see that in both cases we recover the same extension and morphology of the radial profile.
    If no diffuse emission was present on the larger scale, we would have expected to observe patches of localised low surface brightness emission from un-subtracted sources. \par
    In addition to this test, we produced mock LOFAR observations to further investigate the possibility that a blend of point sources at low resolutions can form an outer component around the RH, as suggested in \cite{Rajpurohit2025}. In the simulated observation, we injected a RH similar to the one observed in A2244. We divided two cases, in the first one we injected the real distribution of sources obtained from the two MeerKAT images, while in the second case we injected a uniform distribution of point sources assuming a constant flux. An in depth explanation of this procedure is given in Appendix \ref{appendix:mock_obs}. In both cases we observe a pure exponential profile, with only a small deviation from the injected profile when not masking evident residuals in the first case. The resulting images and radial profiles are shown in Appendix \ref{appendix:mock_obs}. These results suggest that the observed outer component is not related to un-subtracted point sources but is related to the diffuse emission present in the cluster.

\subsection{Radio-X-ray correlation}\label{ptp-corr-sect}
\begin{figure}
        \centering
        \includegraphics[width = 0.85\linewidth]{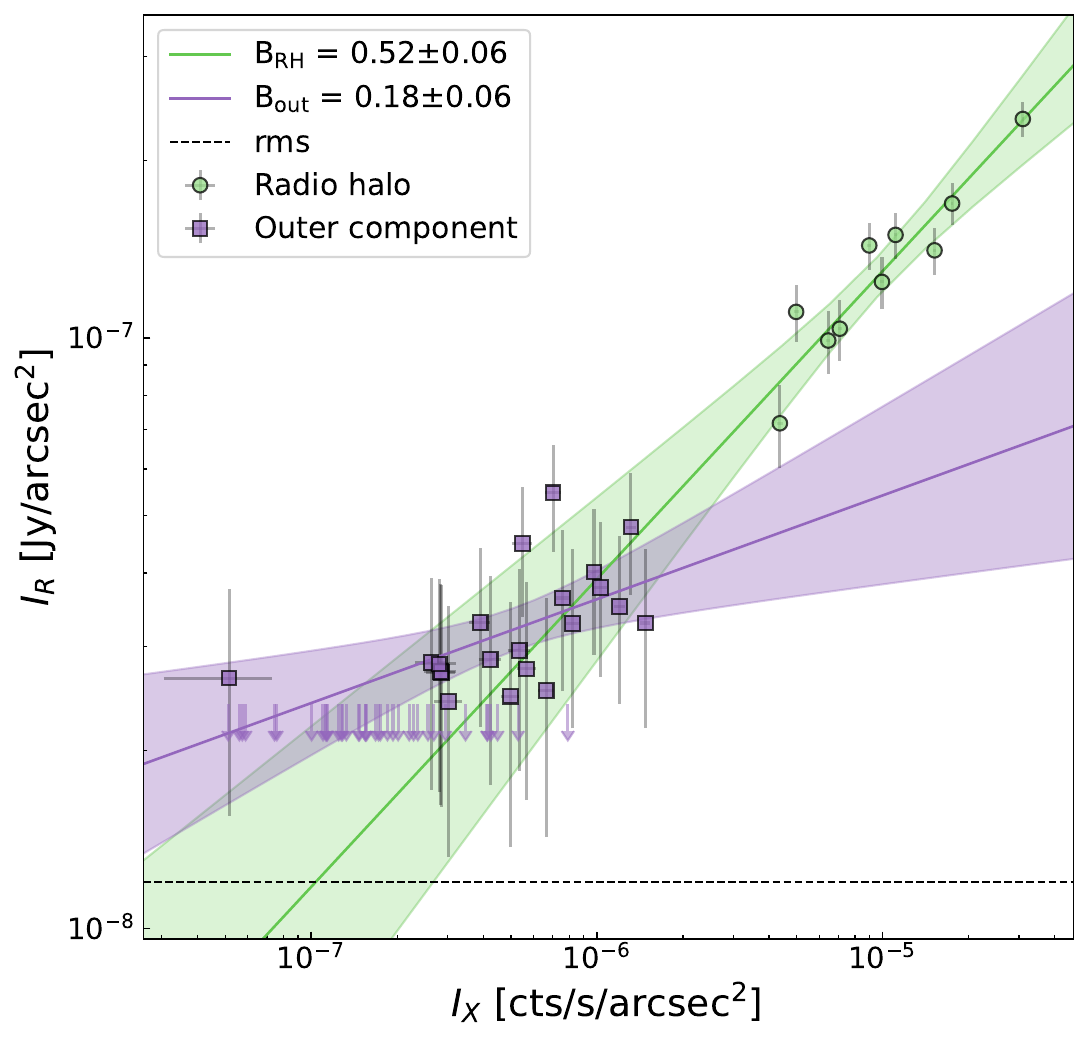}
        \caption{Radio-X-ray surface brightness correlation at $813~\mathrm{MHz}$. We show the points related to the RH emission in green and the points from the outer emission in purple. The green line is the best-fit line for the RH correlation and in purple the best-fit relation for the outer emission. The bands show the $95\%$ confidence region of the regression lines. The rms is $0.012~\mathrm{\mu Jy/arcsec^2}$. The upper limits correspond to the cells where $I_R < 2\sigma_{rms}$ and are not used to search and fit the correlation.}
        \label{fig:ptp-corr}
        \vspace{-3mm}
    \end{figure}

    After deriving the surface brightness of the radio and X-ray emissions in the cells spanning the area within $r_{500}$ and dividing the points between the ones belonging to the RH and the ones that are part of the outer component, we fitted them with the linear relation in Eq. \ref{ptprelation}.
    The results from this analysis are shown in Fig. \ref{fig:ptp-corr} at $813~\mathrm{MHz}$ and in Fig. \ref{fig:ptp-lband} at $1279~\mathrm{MHz}$, where in green we show the points from the inner emission and in red the points from the outer component.
    We exclude all the cells in which $I_R < 2 \sigma_{rms}$ and plot them as upper limits in Figs. \ref{fig:ptp-corr} and \ref{fig:ptp-lband}, although we did not use them to fit the correlation.
    From Fig. \ref{fig:ptp-corr}, we see that the points that are part of the RH emission are well fitted with a sub-linear relation with slope $B_{RH} = 0.52 \pm 0.06$, which is well consistent with the results from \cite{Balboni2024} at $144~\mathrm{MHz}$.
    We confirm the presence of such a relation through the Spearman $r_s$ coefficient, with a value of $r_s = 0.9$ and a p-value of $p = 3.4 \cdot 10^{-4}$.
    At $813~\mathrm{MHz}$ we see that the outer emission follows a correlation with the ICM emission, with a slope of $B_{out} = 0.18 \pm 0.06$.
    The Spearman $r_s$ coefficient of $r_s = 0.64$ points towards a moderate relation, confirmed by $p = 1.7 \cdot 10^{-3}$.
    At higher frequencies, the correlation for the inner emission shows $B_{RH} = 0.66 \pm 0.13$, while for the outer component we find $B_{out} = 0.11 \pm 0.03$, with $r_s = 0.76$ and a low p-value of $p = 0.028$, which is not statistically significant to confirm the presence of the correlation.
    The large number of points below the $2\sigma_{rms}$ , combined with the fact that we are not using them for the fit, can induce a bias in the slope of the correlation for the outer component; in fact, only using the points above the threshold might induce a flattening in the final result. This is also pointed out by the large uncertainties in the points of the outer component. More sensitive observations will help  us to detect fainter regions of the emission and put stronger constraints on this correlation, or possibly show that this correlation is consistent with the RH one. \par
    In Fig. \ref{appendix_fig:grids} we show the grids used to produce the point-to-point correlations. 
    The cells are colour coded following the same scheme of the correlation plots.
    
\section{Discussion}\label{discussion}
\begin{figure}
        \centering
        \includegraphics[width=0.85\linewidth]{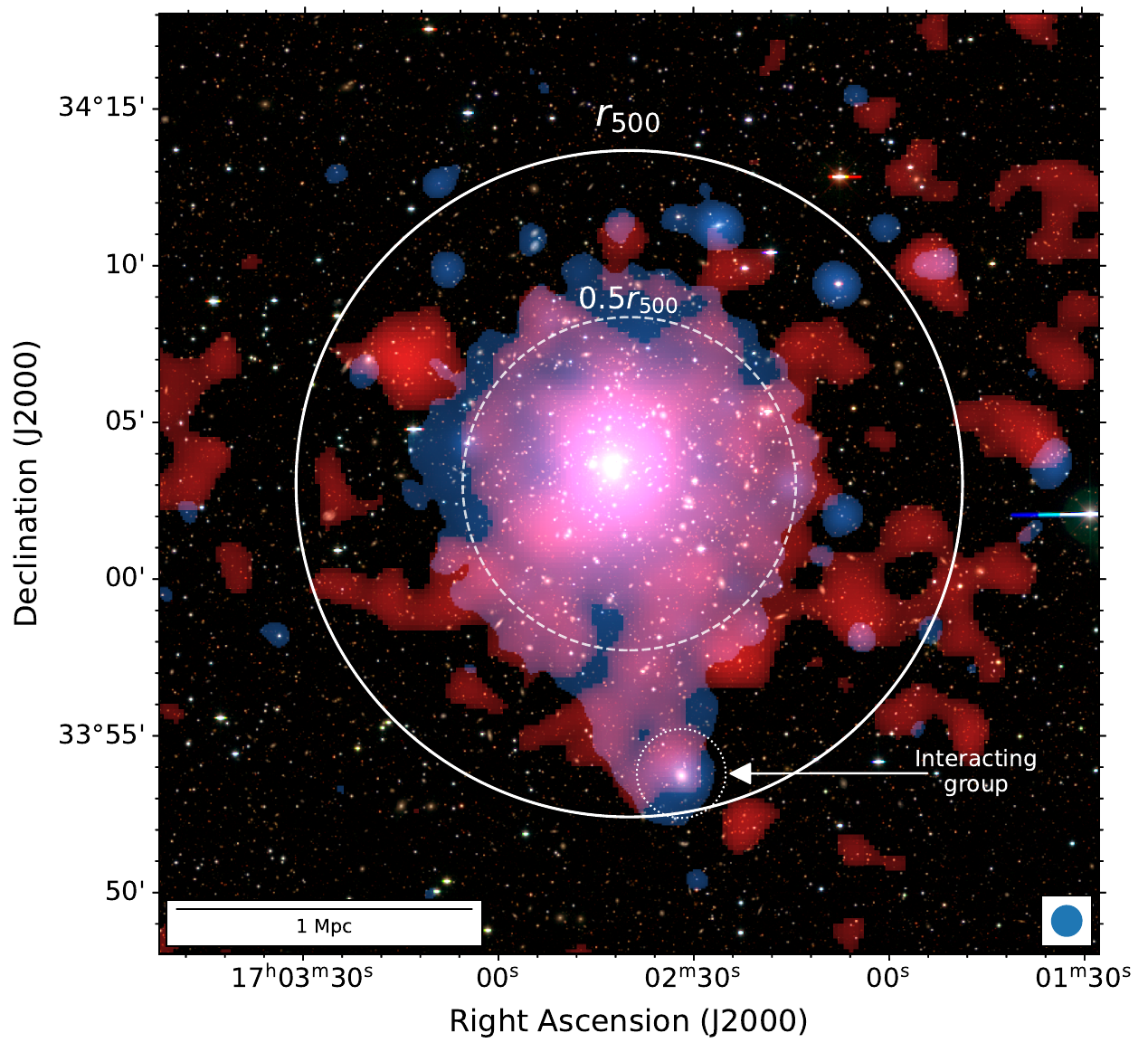}
        \caption{Optical (background, DESI), X-ray (blue, \textit{XMM-Newton}), and radio (MeerKAT UHF) composite image of A2244.}
        \label{fig:rgb_a2244}
        \vspace{-3mm}
    \end{figure}

    The presence of the RH in A2244 can be explained by the minor merger with the nearby group, as it is injecting turbulence in the ICM which can re-accelerate fossil CRe and produce diffuse radio emission, even in the outskirts of the cluster.
    The hypothesis that the origin of the diffuse emission is related to the group interaction is supported by the strong similarities in morphology between the radio and X-ray diffuse emission.
    We see from Fig. \ref{fig:rgb_a2244} that the radio emission is almost co-spatial with the X-ray one, as the outer component extends towards the group, following the X-ray bridge, while part of it extends perpendicularly to it.
    Due to the low mass of the interacting group and of the type of interaction between the two objects, the fraction of gravitational energy transferred to the turbulence is small, meaning that the acceleration of CRe should be inefficient, from which we expect ultra-steep spectrum \citep[$\alpha > 1.6$;][]{Cassano2010} diffuse emission even at low frequencies \citep[an example of this has been observed in Abell 3562 by][]{Venturi2022}.
    The RH in A2244 does not have the spectral properties of emission originated from very low efficiency acceleration mechanisms, i.e. an ultra-steep spectrum. The spectral index is flatter than the average one of RHs related to major merger events \citep[$\alpha = 1.3$;][]{Feretti2012}, although a spectral steepening at higher frequencies is observed, which is consistent with the average value of RHs.
    This is not the first RH observed with these properties in a relaxed cluster, \cite{Kale2016} measured the spectral index of the RH in the cluster CL1821+643, which experienced an off-axis merger that left the cool-core intact, obtaining a value of $\alpha = 1.0 \pm 0.1$, while \cite{vanWeeren2026} discovered a mH+RH system in Abell 1795, in which the RH shows a spectral index of $\alpha = 1.1 \pm 0.1$.
    Recent studies \citep{Lusetti2024, Pignataro2024, vanWeeren2024, Groeneveld2026} are showing that minor mergers can inject enough turbulence to power megaparsec-scale radio emission, although this has been observed in systems hosting a mH and a RH.
    Some works have also shown that when a cluster hosts a mH and a RH, the latter has a steeper spectral index due to less energy being available in the outer region \citep[e.g.][]{Biava2021}.
    In general we see from the observations that minor mergers can generate RHs without disrupting the cool-core, but it is not clear what drives the steepness of the spectrum of RHs around mHs. \par
    In Fig. \ref{fig:PM} we show the computed radio power of the RH in A2244 (in red) alongside with the luminosities of the RHs of the \textit{Planck} clusters of the LoTSS-DR2 \citep{Botteon2022a, Cuciti2023}. 
    From the radio power-mass relation we show that the RH is under-luminous with respect to the linear correlation found by \cite{Cuciti2023}. In their study, \cite{Cuciti2023}, also observed that clusters that are less dynamically disturbed host RHs that are under-luminous compared to the P-M relation. Our results for the RH in A2244 are consistent with this result.
    We also did not take into account the luminosity of the outer component, as the flux density has not been estimated analytically and combining the two measures would be unreliable. \par
    We investigated the possibility that the diffuse emission in this cluster is a combination of a central mH enveloped by a RH.
    As the inner component of this system shows morphological properties intermediate between a mH and a RH, the only approach we have to classify it is by using the point-to-point correlation between $I_R$ and $I_X$, since for mHs this correlation is super-linear \citep{Ignesti2020} and for RHs is sub-linear \citep{Giacintucci2005, Hoang2021, Balboni2024}.
    In Sect. \ref{ptp-corr-sect}, we showed how the outer component of the emission follows a shallow sub-linear relation observed at $813~\mathrm{MHz}$, with a slope of $B = 0.18 \pm 0.06$, while the inner component of the emission follows a steeper sub-linear relation with $B = 0.52 \pm 0.06$.
    The results of the point-to-point correlation for the inner component are consistent with the results from \cite{Balboni2024} at $144~\mathrm{MHz}$.
    As the analysis shows a sub-linear trend for both the inner and the outer component, we tend to exclude the possibility that the system consists in a mH enveloped by a RH.
    Another possibility is that the outer component is a radio bridge \citep{Govoni2019}, which has been observed between pairs of clusters \citep[e.g.;][]{Govoni2019, Botteon2020b}, and for which the emission is concentrated between the two objects.
    These bridges have been also observed between cluster-group pairs, such as in the Coma cluster \citep{Bonafede2021, Bonafede2022} and in Abell 3562 \citep{Venturi2022}.
    Another case of peculiar radio emission has been observed in the cluster Abell 2061 by \cite{Pignataro2024}. This cluster shows a RH with an extension towards the neighbouring cluster Abell 2067, although it is not clear whether this emission is a radio bridge or an extension of the RH.
    In the case of A2244, the outer component is not present in the southern region only, on top of the X-ray bridge, but it is also found in the E-W direction, as we can see from Fig. \ref{fig:rgb_a2244}.
    Despite this, we see that the outer component follows a point-to-point radio X-ray correlation similar to the ones observed by \cite{Botteon2020b} and \cite{deJong2022} in radio bridges.
    We cannot completely exclude the possibility that the observed outer component is a bridge between A2244 and the interacting group.
    Nonetheless, considering the morphology of the diffuse radio emission, we lean towards the possibility of this being either a RH that has been disturbed by the passage of the group or another diffuse radio component, such as an intra-cluster bridge.
    The simulation from \cite{Beduzzi2023, Beduzzi2024}, presents a similar scenario to the one observed in this cluster, showing that groups interacting with the cluster can drag CRe on their path, forming an outer and highly asymmetrical component in the diffuse emission, while also injecting turbulence in the outskirts of the cluster, which could re-accelerate fossil CRe in these regions. In the simulation these sources appears similar to megaH.
    If this is the case for A2244, this scenario could explain the morphology of the diffuse radio emission, and the possibility that the sources are formed from one single merging interaction.
    This could also lead to the presence of a moderate surface brightness relation between the outer component and the ICM emission alongside with the one between the ICM and the RH. \par
    We now discuss similarities and differences with respect to a megaH interpretation.
    The outer component of the emission shows some similarities with the megaHs reported in \cite{Cuciti2022}, as it has a highly asymmetric morphology and a non-exponential surface brightness profile. The main difference regards the spectral index of the outer component when compared with the ones measured for the two megaHs in \cite{Cuciti2022}.
    In fact, its integrated spectral index is flatter with respect to the ones measured for the megaHs in Abell 665 and ZwCl 0634.1+4750, where from a study at $54~\mathrm{MHz}$ and $144~\mathrm{MHz}$ the two megaHs show a spectral index of $\alpha \approx 1.6$ \citep{Cuciti2022}, while in A2244 we observe $\alpha^{1279}_{144} = 0.80 \pm 0.04$.
    Although our spectral index is flatter compared to those measured in megaH, we must underline that the spectral index measurements in \cite{Cuciti2022} have been obtained by small regions drawn on the images and can be biased towards steeper indices.
    Taking into account similarities and differences from the known megaHs, the main hypothesis for the classification of this emission is a highly disturbed RH. However, we leave open the speculative suggestion that the outer component a megaH, as the shape of the radial profile is similar to what has been observed for this kind of sources \citep{Cuciti2022}. For this it is important to increase the statistics about megaHs to better define their properties. Deeper radio and X-ray observations will help us constrain better the nature of this diffuse radio emission. \par
    Works in the literature show us that minor mergers can inject enough turbulence to accelerate particles to high frequencies and form large-scale diffuse radio emission \citep[][]{Kale2016, Biava2021, Biava2024, vanWeeren2026, Groeneveld2026}.
    The main hypothesis is that the origin of the emission in A2244 is strictly related to the passage of the nearby group, which is injecting turbulence, especially in the southern part of the cluster.
    Current theoretical models cannot explain the flat spectrum  of the two components, hence we refrain to give possible explanations for our finding. \par
    The complexity of the emission in this cluster shows us that the general picture of a sharp classification of diffuse cluster sources has become increasingly blurred with the discovery of emission with properties that are mixed within different scenarios and classes, which has been the trend over the past few years \citep{Cuciti2022, Botteon2022b, Bruno2023a, Biava2024, vanWeeren2024}.
    To conclude, the discovery of such faint diffuse emission is expanding our knowledge on the physical processes that are taking place in clusters, and with these observations we are pushing the limits of what is achievable with the current instruments.
    With the advent of the Square Kilometre Array \cite[SKA;][]{Dewdney2009}, we will be able to detect even the faintest of these sources, obtaining even higher precision in the flux density and spectral index measurements, thanks to its increase in sensitivity, with a factor of almost ten, and to its high resolution ($2.5\arcsec$ in Band 1 and $1\arcsec$ in Band 2).
    Despite its lower resolution compared with LOFAR, SKA-Low will also help us to study and detect more of these sources, providing an increase in sensitivity by a factor of eight, which will allow us to improve the statistics of these sources, even in the faintest end.

\section{Conclusions}\label{conclusions}
    In this paper we presented a multi-frequency study of the galaxy cluster Abell 2244 using LOFAR HBA, MeerKAT UHF, and L-band observations.
    The cluster hosts a low luminosity RH that was reported for the first time by \cite{Botteon2022a}, while \cite{Balboni2024} carried out a first radio-X-ray analysis of the diffuse emission, also claiming the detection of a two-component emission. 
    We summarize the conclusions of our study:
\begin{itemize}
    \item We detected the RH at higher frequencies and estimated its spectral index, finding an integrated value of $\alpha^{1279}_{144} = 0.9 \pm 0.1$ and spectral steepening between $813~\mathrm{MHz}$ and $1279~\mathrm{MHz}$. Due to the size of the region in which we measured the spectral index and the fact that we measured the flux densities from the image, the spectral index is only characteristic of the central region of the RH and might be contaminated by the fainter emission from the second component. The RH is under-luminous with respect to the radio power-mass relation found by \cite{Cuciti2023} and its spectral index is comparable with those of RHs found in disturbed clusters; this points towards a complex formation scenario, as the cluster is not strongly disturbed. Using the observations at $813~\mathrm{MHz}$ and $1279~\mathrm{MHz,}$ we confirm the presence of diffuse radio emission and the results of the point-to-point relation found by \cite{Balboni2024} at $144~\mathrm{MHz}$.    \item We also detected the outer component in the cluster in the MeerKAT observations. This component shares a similar morphology with the X-ray diffuse emission, which means that it is possible that its origin is related to the cluster-group interaction. The integrated spectral index of the outer component is $\alpha^{1279}_{144} = 0.80 \pm 0.04$, which is consistent with the value found for the RH, although in this case we do not observe any high frequency spectral steepening. At $813~\mathrm{MHz}$ we observe a radio X-ray correlation between the outer component and the ICM emission, with a slope shallower than the one observed for the RH, although there are a large number of upper limits in the correlation that might bias the result. This result must be confirmed with deeper and more sensitive observations. This correlation is not observed at $1279~\mathrm{MHz}$, possibly due to the lower sensitivity of the observations to the diffuse low brightness emission.
    \item From the results of the $I_R-I_X$ correlation, we can exclude the possibility that the diffuse emission observed is a combination of a mH surrounded by a RH. By analysing the physical and morphological properties of the diffuse radio emission in the cluster, we cannot give a classification of the diffuse emission. This is related to the complexity in clearly defining the properties of the outer component, which shows properties comparable both with RHs and megaHs. From this, we speculate that the source could be a RH that has been disturbed by the minor merger, elongating its morphology in the direction of the merger, which is highlighted by the X-ray bridge, or a megaH, as the radial profile and its extension is comparable to that of observed megaHs.
    \item We investigated the possibility that the outer component in A2244 is formed by faint un-subtracted sources blended at low resolutions and is not real diffuse emission, as suggested by \cite{Rajpurohit2025}. We created a mock LOFAR observation, in which we injected a RH, then, in two different runs, we also injected different point source distributions, dividing them based on the distribution and flux density of the sources. While in both simulations we do not observe the formation of an outer component around the RH, we see that, in the first case, badly subtracted sources can slightly contribute to the diffuse radio emission in its peripheral regions, where the brightness is lower. Overall, this contribution is still not enough to explain the presence of the outer component in A2244. These mock observations do not take into account discrete extended sources as the only one observed in the cluster is not in the regions where the outer component of the diffuse emission is observed.
\end{itemize}

\begin{acknowledgements}
M.C., F.D.G., C.G., J.M.B., M.D.C. and G.D.G. acknowledge support from the ERC Consolidator Grant ULU 101086378. 
A.B. and M.B. acknowledge financial support from the ERC CoG BELOVED n. 101169773.
F.G. and M.B. acknowledge the financial contribution from the INAF GO grant 1.05.24.02.10 Extended Radio Emission in Galaxy Clusters at deep focus with MeerKAT. 
G.B. acknowledges financial contribution from INAF Theory grant 1.05.23.06.01 Theory and simulations of non-thermal phenomena in galaxy clusters and beyond.
This research made use of the LOFAR-IT computing infrastructure supported and operated by INAF, including the resources within the PLEIADI special ``LOFAR'' project by USC-C of INAF, and by the Physics Dept. of Turin University (under the agreement with Consorzio Interuniversitario per la Fisica Spaziale) at the C3S Supercomputing Centre, Italy. 
LOFAR \citep{vanHaarlem2013} is the Low Frequency Array designed and constructed by ASTRON. It has observing, data processing, and data storage facilities in several countries, which are owned by various parties (each with their own funding sources), and that are collectively operated by the ILT foundation under a joint scientific policy. The ILT resources have benefited from the following recent major funding sources: CNRS-INSU, Observatoire de Paris and Université d'Orléans, France; BMBF, MIWF-NRW, MPG, Germany; Science Foundation Ireland (SFI), Department of Business, Enterprise and Innovation (DBEI), Ireland; NWO, The Netherlands; The Science and Technology Facilities Council, UK; Ministry of Science and Higher Education, Poland; The Istituto Nazionale di Astrofisica (INAF), Italy. This research made use of the Dutch national e-infrastructure with support of the SURF Cooperative (e-infra 180169) and the LOFAR e-infra group. The Jülich LOFAR Long Term Archive and the German LOFAR network are both coordinated and operated by the Jülich Supercomputing Centre (JSC), and computing resources on the supercomputer JUWELS at JSC were provided by the Gauss Centre for Supercomputing e.V. (grant CHTB00) through the John von Neumann Institute for Computing (NIC). This research made use of the University of Hertfordshire high-performance computing facility and the LOFAR-UK computing facility located at the University of Hertfordshire and supported by STFC [ST/P000096/1], and of the Italian LOFAR IT computing infrastructure supported and operated by INAF, and by the Physics Department of Turin university (under an agreement with Consorzio Interuniversitario per la Fisica Spaziale) at the C3S Supercomputing Centre, Italy.
The MeerKAT telescope is operated by the South African Radio Astronomy Observatory, which is a facility of the National Research Foundation, an agency of the Department of Science, Technology and Innovation.
We acknowledge the developers of the following \textsc{Python} packages which were used in this work: \textsc{AstroPy} \citep{astropy:2013, astropy:2018, astropy:2022}, \textsc{Matplotlib} \citep{matplotlib}, \textsc{SciPy} \citep{SciPy-NMeth}, \textsc{APLpy} \citep{aplpy2012, aplpy2019}, \textsc{NumPy} \citep{numpy} and \textsc{PANDAS} \citep{pandas-mckinney, pandas}. 

\end{acknowledgements}

\bibliographystyle{aa}
\bibliography{bibliography}

\onecolumn
\begin{appendix}
\section{Intermediate-resolution images}\label{appendix_20asec}
\begin{figure}[!htb]
    \centering
    \includegraphics[width=0.34\textwidth]{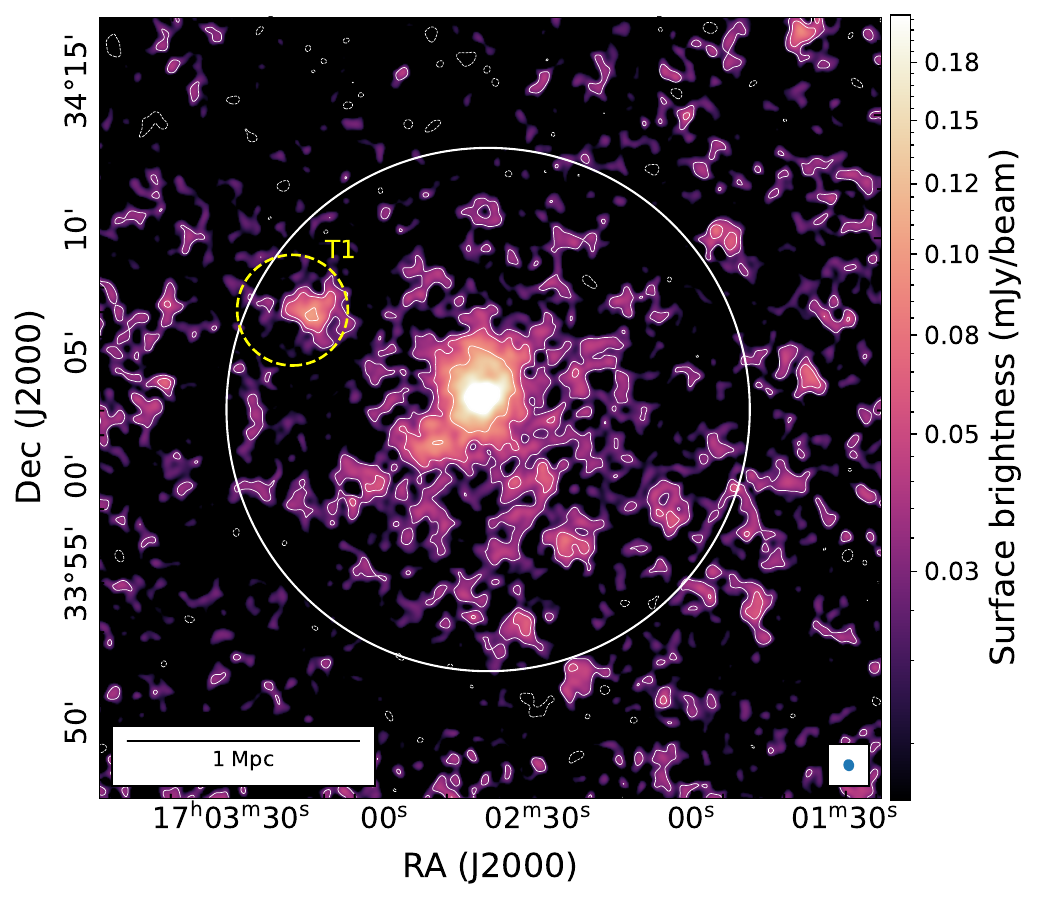}
    \includegraphics[width=0.34\textwidth]{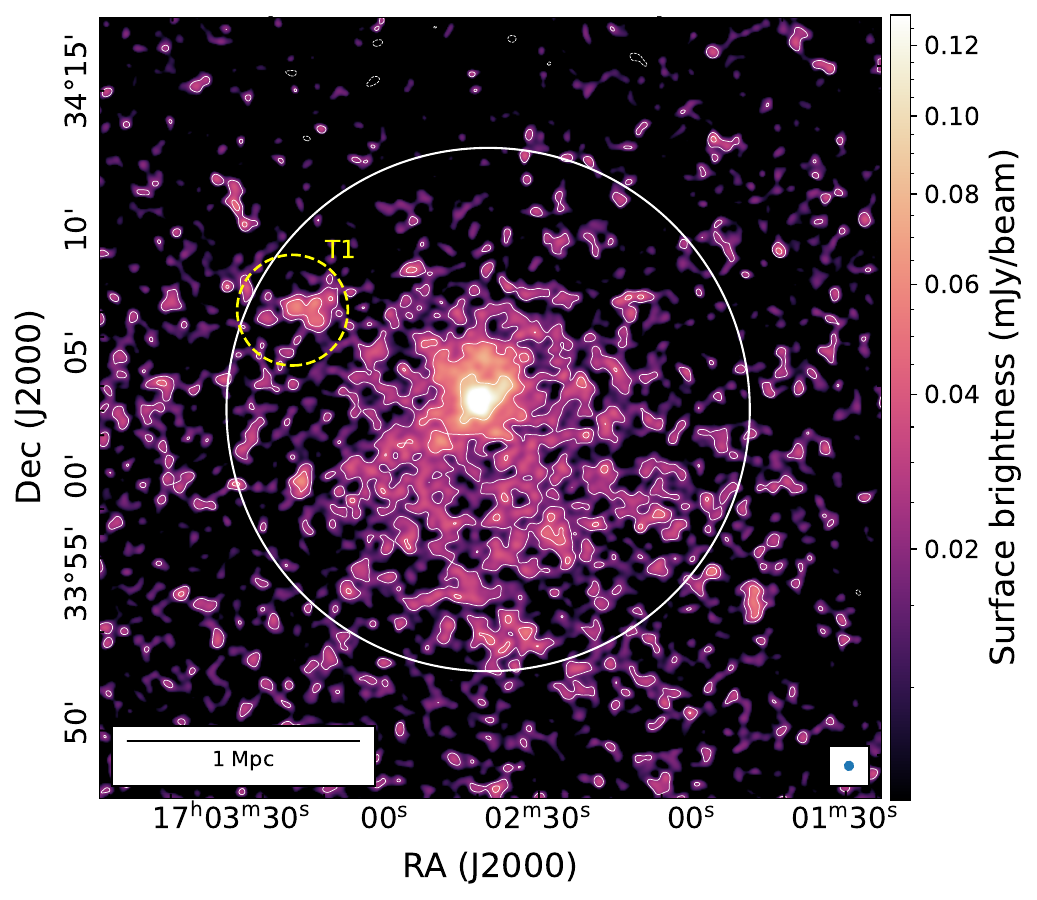}
    \includegraphics[width=0.34\textwidth]{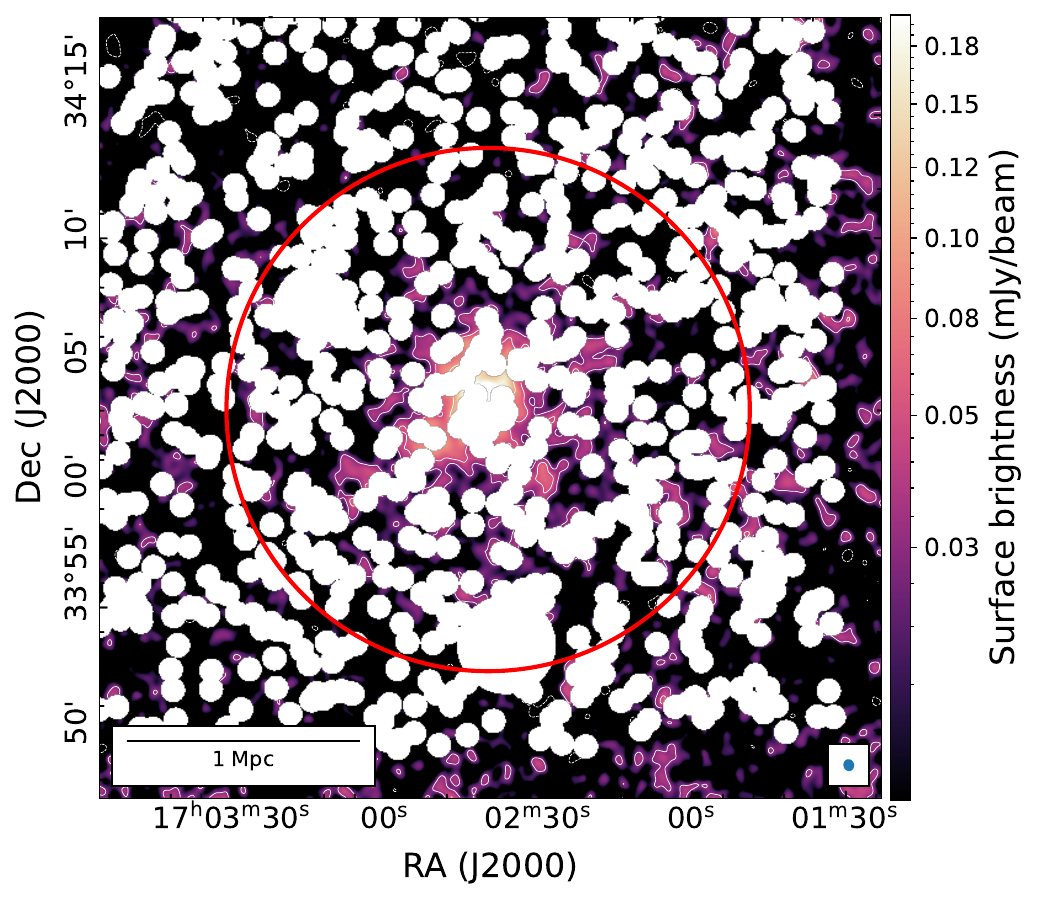}
    \includegraphics[width=0.34\textwidth]{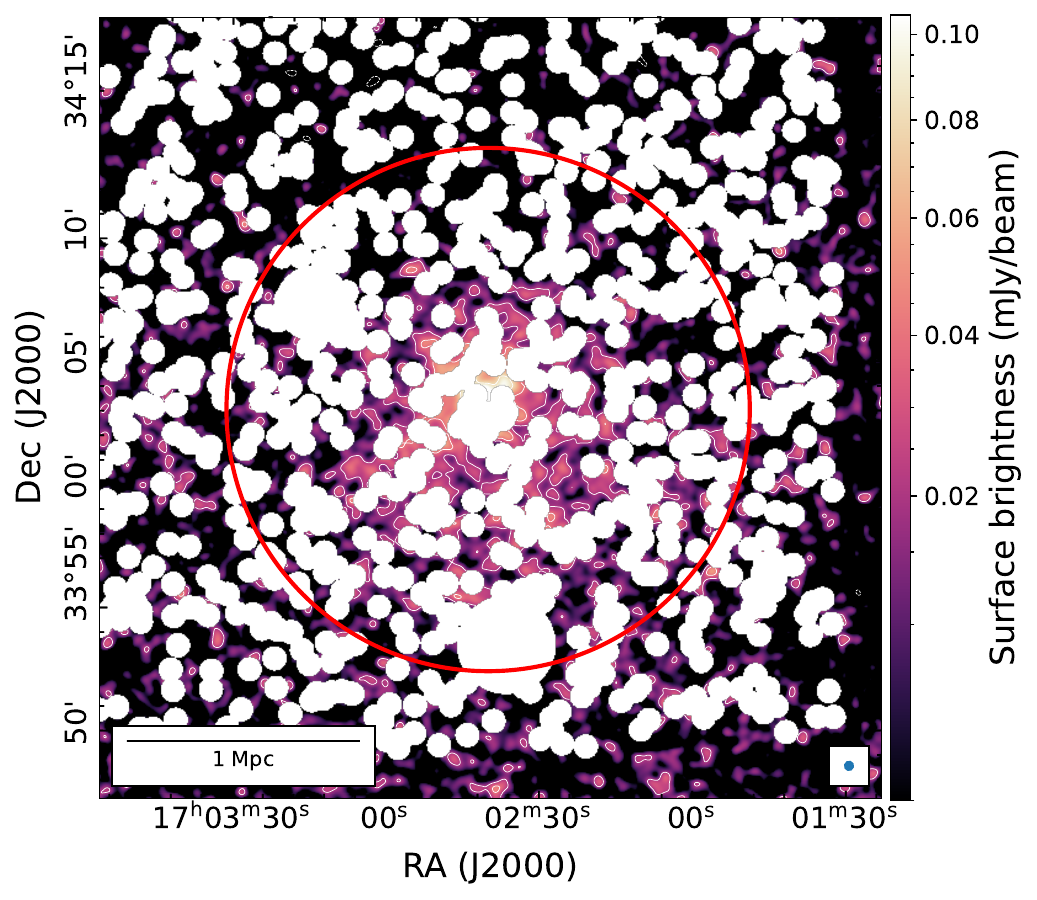}
    \caption{UHF (\textit{left}) and L (\textit{right}) bands $25\arcsec$ source-subtracted images of A2244. The white circle (top images) and red circle (bottom images) is $r_{500}$. For MeerKAT UHF the noise is $\sigma_{rms} = 12 \ \mathrm{\mu Jy/arcsec^2}$, while for MeerKAT L-band is $\sigma_{rms} = 10 \ \mathrm{\mu Jy/arcsec^2}$. The contours are set at levels $[-3, 2, 4, 8] \times \sigma_{rms}$. The images at the top are without masking the location of the point sources. In the top images we circled the location of T1 in yellow.}
    \label{fig:r500_20asec}
\end{figure}

In this section we show the $25\arcsec$ resolution source-subtracted images in UHF and L-bands.
We masked all the positions of sources identified in L-band and UHF at $5\sigma_{rms}$ and cross-matched the two catalogues, including sources only detected in one of the two images, to be more conservative when masking, to produce the profile in Fig. \ref{fig:sb_comparison}.
We see strong residual emission in the location of T1, which we highlighted in Fig. \ref{fig:r500_20asec}.
This emission has been masked when making all the surface brightness profiles showed in this paper.
\FloatBarrier

\section{Radio halo and outer component flux density regions}
In this section we show the regions used to measure the flux densities of the RH and of the outer component in the three images.
\begin{figure}[!htb]
    \centering
    \includegraphics[width = 0.3\textwidth]{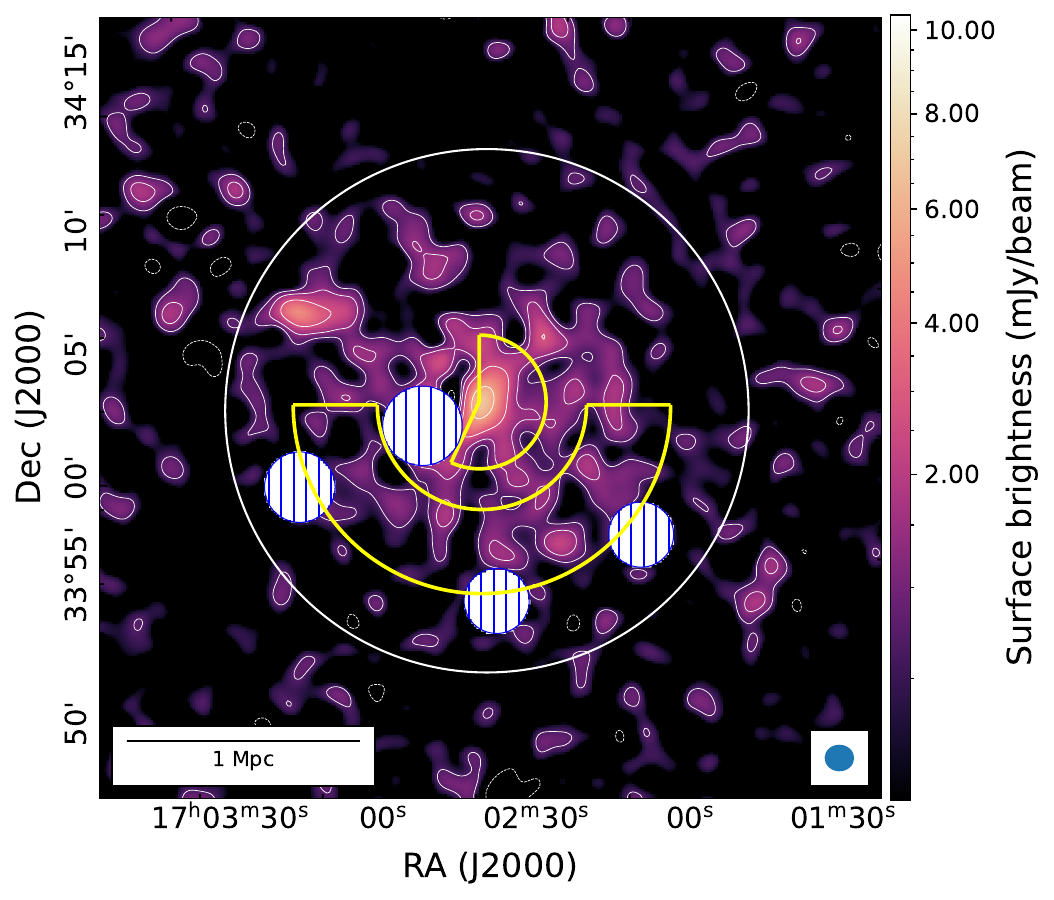}
    \includegraphics[width = 0.3\textwidth]{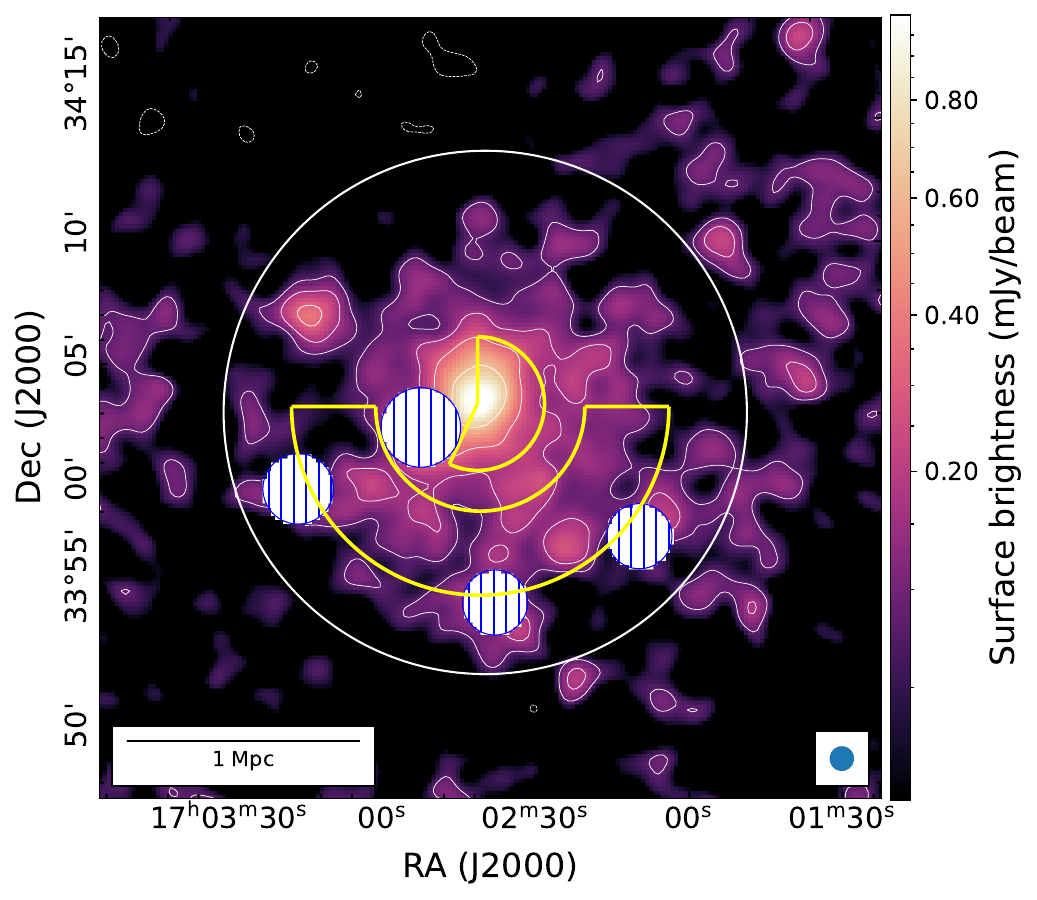}
    \includegraphics[width = 0.3\textwidth]{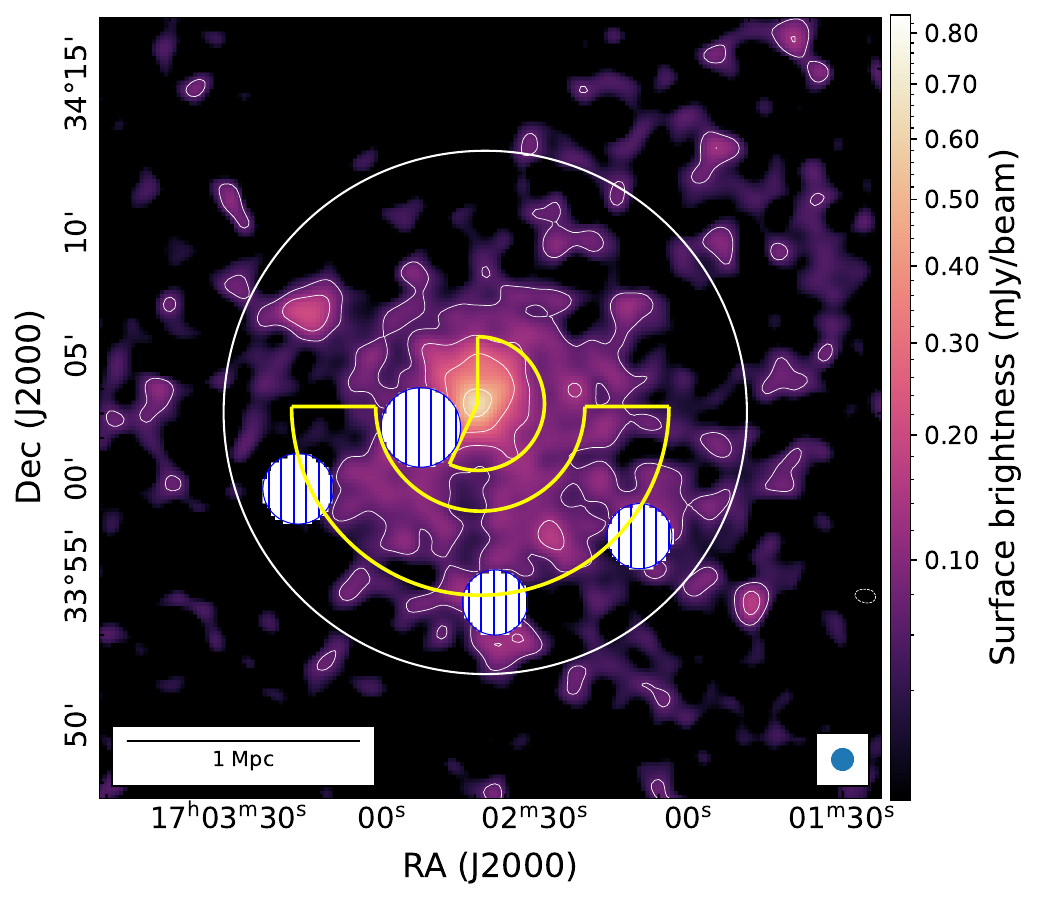}
    \caption{Low-resolution images ($\approx 60\arcsec$) of A2244 at $144~\mathrm{MHz}$ (\textit{left}), $813~\mathrm{MHz}$ (\textit{centre}), and $1279~\mathrm{MHz}$ (\textit{right}). The yellow regions are used to estimate the flux densities of the RH and of the outer component. The white circle is $r_{500}$. We masked regions where subtraction residuals were presents.}
    \label{appendix_fig:flux_regions}
\end{figure}
\FloatBarrier

\section{Point-to-point correlation in MeerKAT L band}
In this section we show the results for the point-to-point radio-X-ray analysis obtained from the MeerKAT L band image.
\begin{figure}[!htb]
    \centering
    \includegraphics[width = 0.4\textwidth]{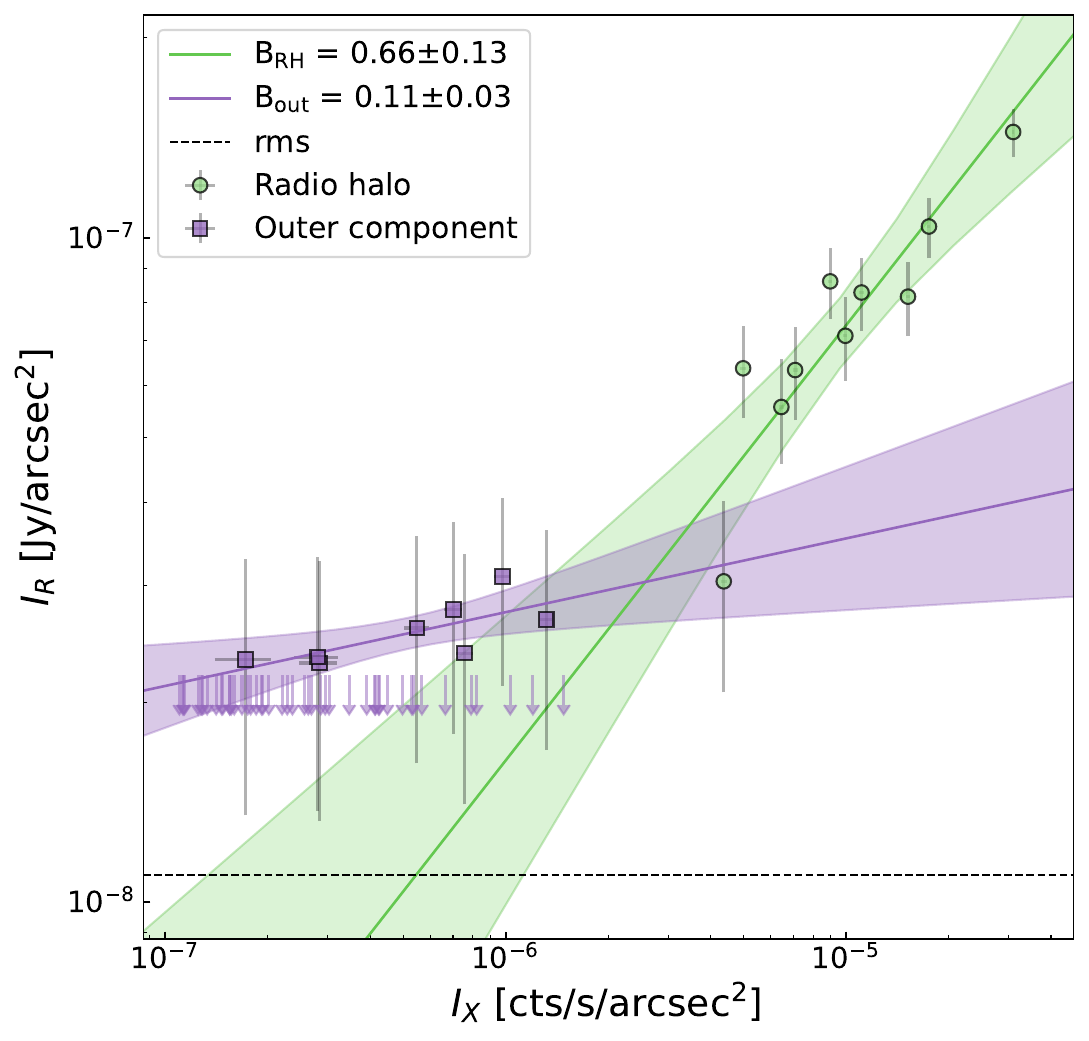}
    \caption{Radio-X-ray surface brightness correlation using the MeerKAT L-band image. We show in green the points related to the RH emission and in purple the points from the outer emission. The green line is the best-fit line for the RH correlation, while the green bands show the $95\%$ confidence region of the regression line, while the purple ones are related to the outer component. The rms is $0.0011~\mathrm{\mu Jy/arcsec^2}$. The upper limits corresponds to the cells where $I_R < 2\sigma_{rms}$ and are not used to search and fit the correlation.}
    \label{fig:ptp-lband}
\end{figure}
\FloatBarrier

\section{Spectral index error map}
In this section we show the spectral index error map related to the spectral index map shown in Fig. \ref{fig:spix_maps}
\begin{figure}[!htb]
    \centering
    \includegraphics[width = 0.4\textwidth]{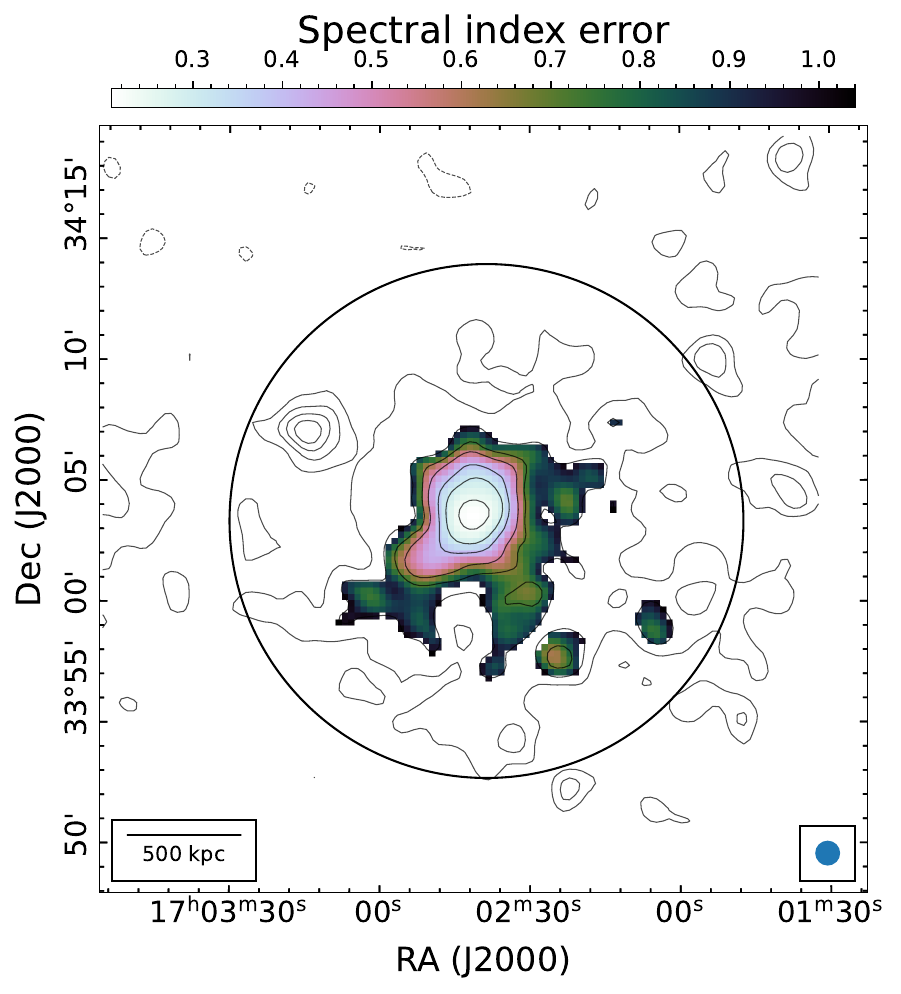}
    \caption{Spectral index error map. The beam size is $61.8\arcsec \times 61.8\arcsec$. The contours are from the UHF low-resolution source-subtracted image and are set at $[2, 4, 8, 16, 32, 64] \times \sigma_{rms}$, where $\sigma_{rms} = 40 \ \mathrm{\mu Jy/beam}$. The black circle is $r_{500}$.}
    \label{fig:spix_errmaps}
\end{figure}
\FloatBarrier

\section{Double exponential fitted profiles}\label{appendix_double_exp}
In this section we show the radial profiles fitted with a double exponential model to test the hypothesis of the emission being a combination of a mH and of a RH.
{\renewcommand{\arraystretch}{1.15}
\begin{table*}[!htb]
    \caption{Results of the fit of the diffuse emission.}
    \centering
    \begin{tabular}{cccccc}
        \hline \hline
        Frequency & $I_0 $ & $r_{e1}$ & $I_1$ & $r_{e2}$ & $\tilde{\chi}^2$ \\
        $\mathrm{[MHz]}$ & $\mathrm{[\mu Jy/arcsec^2]}$ & [kpc] & $\mathrm{[\mu Jy/arcsec^2]}$ & [kpc] &  \\
        \hline
        144 & $20 \pm 205$ & $11 \pm 64$ & $0.7 \pm 0.1$ & $313 \pm 41$ & $1.49$ \\ 
        \hline
        813 & $0.47 \pm 0.14$ & $58 \pm 15$ & $0.10 \pm 0.02$ & $442 \pm 46$ & $2.71$ \\
        \hline
        1279 & $0.49 \pm 0.27$ & $36 \pm 14$ & $0.06 \pm 0.01$ & $461 \pm 39$ & $1.75$ \\
        \hline
    \end{tabular}
    \label{tab:fit_table_double_exp}
\end{table*}
}
\begin{figure}[!htb]
    \centering
    \includegraphics[width=0.33\textwidth]{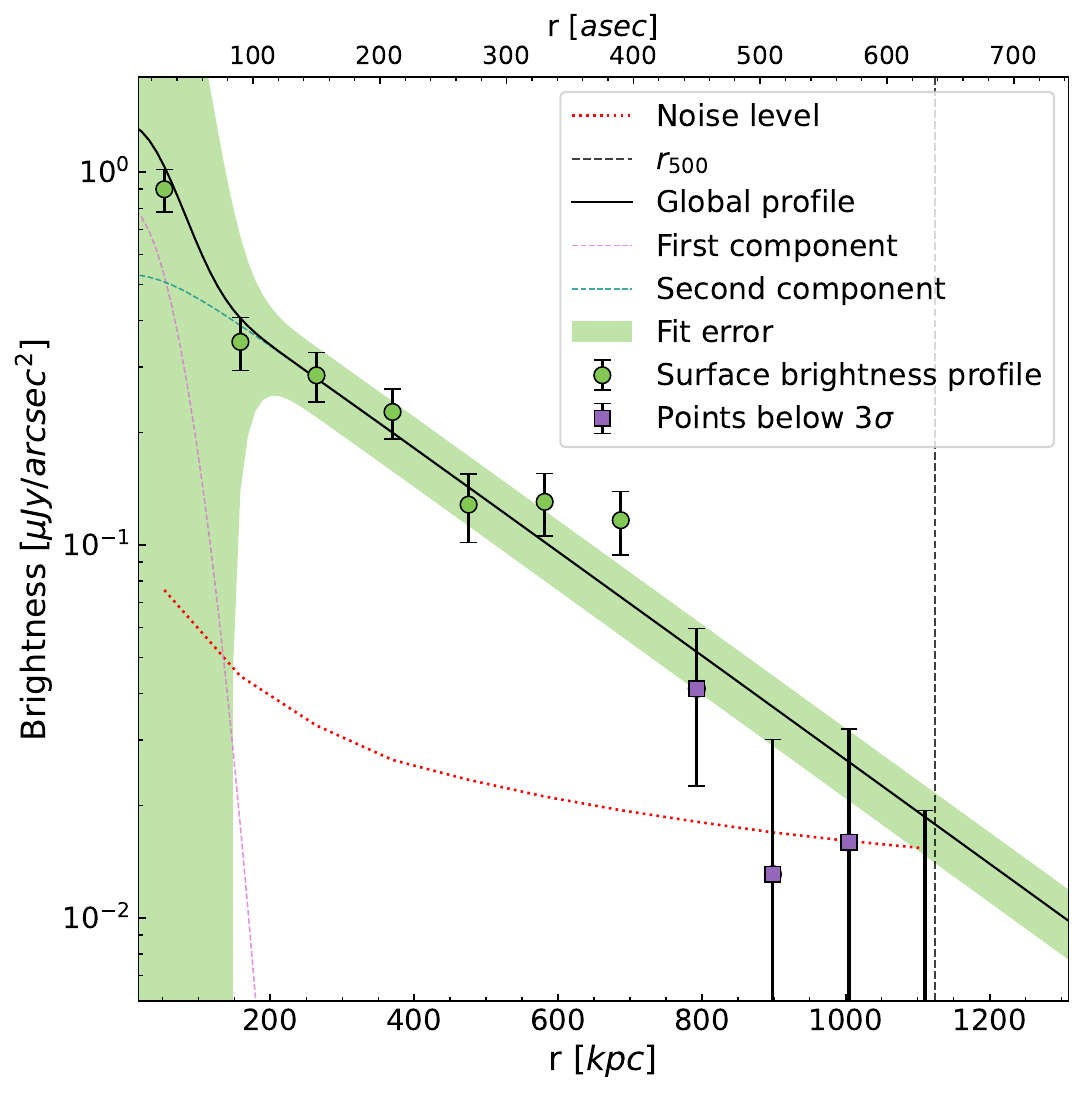}
    \includegraphics[width=0.33\textwidth]{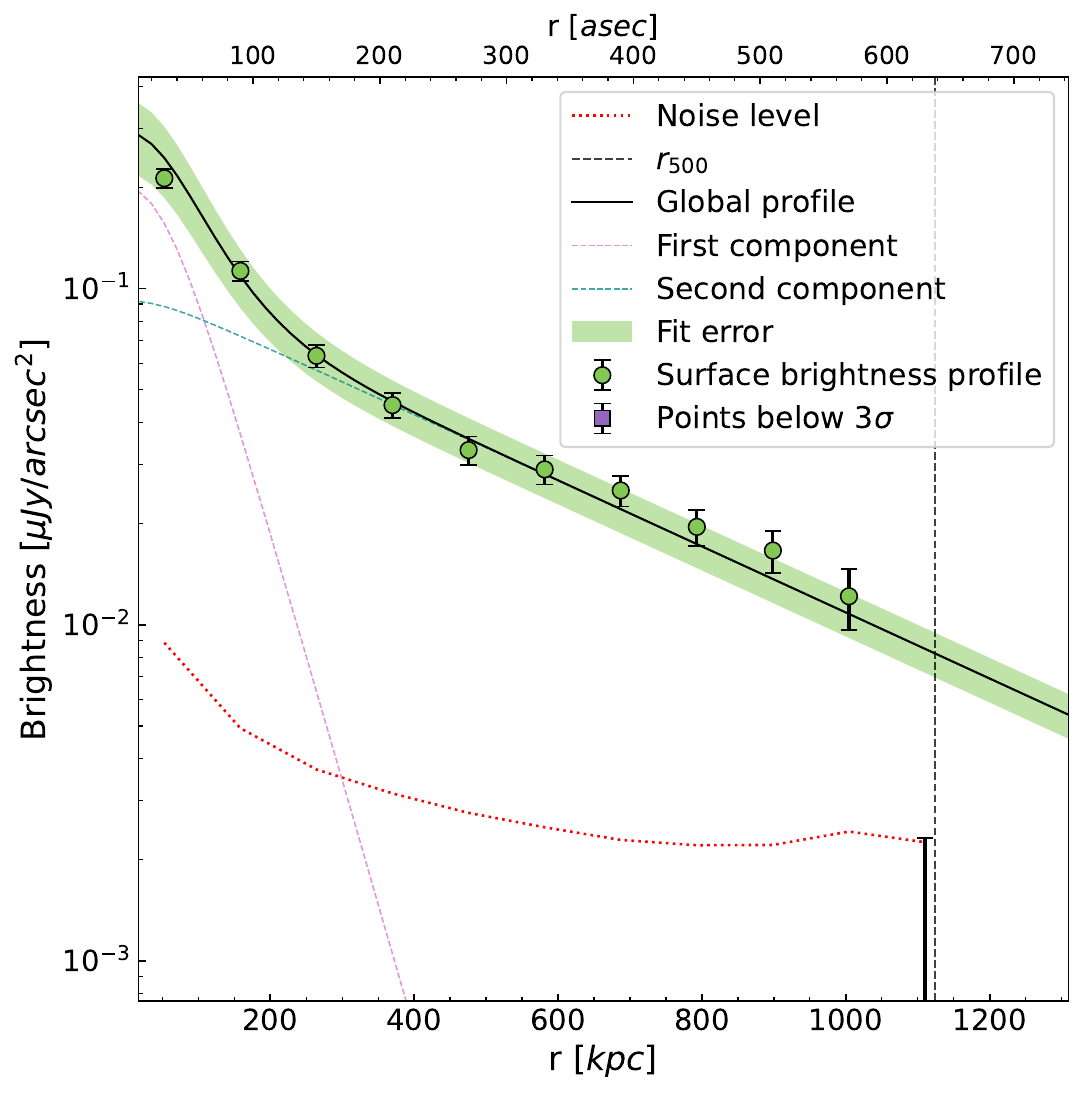}
    \includegraphics[width=0.33\textwidth]{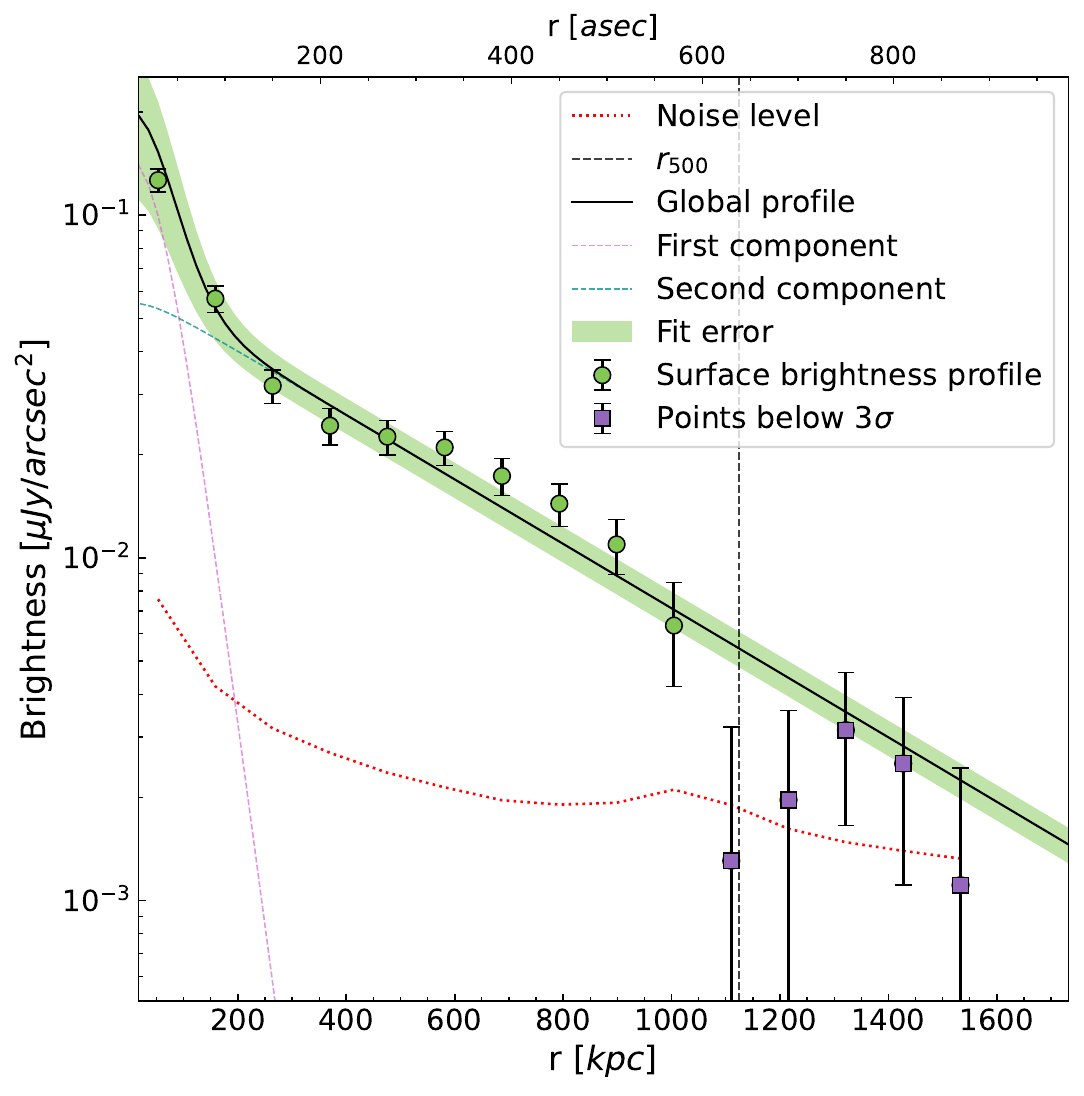}
    \caption{Surface brightness profiles at $144~\mathrm{MHz}$ (\textit{left}), $813~\mathrm{MHz}$ (\textit{middle}), and $1279~\mathrm{MHz}$ (\textit{right}) of the diffuse emission. The solid black line is the fitted profile. The red dotted line is the $1\sigma$ detection limit in each annulus. The vertical dashed line is $r_{500}$.}
    \label{fig:double_exp_prof}
\end{figure}
\FloatBarrier

\section{Mock LOFAR HBA observations}\label{appendix:mock_obs}
    To investigate the origin of the outer component and how faint un-subtracted sources contribute to diffuse emission in clusters, we produced a mock LOFAR HBA observation using the LOFAR Simulation Tool\footnote{\url{https://github.com/darafferty/losito}} \citep[LoSiTo;][]{Edler2021}. We choose to simulate LOFAR observations as they have a lower sensitivity compared to the MeerKAT ones, which allows us to better test the suggestion by \cite{Rajpurohit2025} of faint sources contributing to the diffuse radio emission when their flux densities are comparable to the noise level.
    We simulated an 8 hours long observation, to which we only added Gaussian noise, hence assuming a perfect data calibration, which produced visibilities with $\sigma_{rms,mock} = 107~\mathrm{\mu Jy/arcsec^2}$, similar to the typical noise of high quality LoTSS pointings.
    After adding the noise to our data, we injected in the uv-plane a mock RH having $I_0 = 2.7~\mathrm{\mu Jy/arcsec^2}$ and $r_e = 110~\mathrm{kpc}$, which are the values obtained by fitting the inner component of the emission observed in A2244 at $813~\mathrm{MHz}$ with a single exponential. In Tab. \ref{tab:sim_table}, we report the recovered $I_0$ and $r_e$.
    After this, we injected in the visibilities a distribution of point sources.
    We created the source catalogue from the MeerKAT UHF band image, as it has the best combination of sensitivity and noise levels, using \textsc{PyBDSF} \citep{Mohan2015} and only considering the sources with flux density above $5\sigma_{rms}$, resulting in a total amount of $148$ (projected) sources within $r_{500}$.
    While from the MeerKAT images we detect $148$ sources, in LOFAR we detect only $52$ sources at a $5\sigma$ level, resulting in almost $100$ sources under the noise level in the real LOFAR observation.
    To inject these $148$ sources in the mock observation, we extrapolated their flux density at $144~\mathrm{MHz}$ by determining their spectral index between UHF and L-band and using this value to scale their flux density at lower frequencies.
    As the mock observation and the real LOFAR observation have different noise levels ($\sigma_{rms,obs} = 120~\mathrm{\mu Jy/beam}$ for the real observation and $\sigma_{rms,mock} = 107~\mathrm{\mu Jy/beam}$ for the mock image), to properly simulate the observation, we rescaled the flux density of the point sources by $\sigma_{rms,mock}/\sigma_{rms,obs}$ to maintain the signal-to-noise ratio for each point source.
    This was done so we could reproduce the population of faint sources below the noise of the image, to study their contribution at low resolutions. 
    In the following sections, we refer to this source injection as `Case 1'.
    We followed the same data reduction and analysis method of the real observations (Sect. \ref{source_sub}) to produce the low-resolution source-subtracted images and surface brightness radial profiles. 
    The higher sensitivity of the mock observation compared to the real LOFAR observation is related to the fact that we only simulated the noise, hence obtaining a perfect calibration. This leads to a higher signal-to-noise ratio detection for the RH. If the un-subtracted sources are showing up as an outer component around the RH in the simulated observation, we expect it to have a lower surface brightness compared to the one from the real observation and the appearance of the outer component should be at a greater radii. This is due to the lower relative contribution from the injected sources to the diffuse emission, as they are rescaled based on the ratio of the $\sigma_{rms}$ of the two datasets and thus have a lower flux density. \par
    In addition to the first injection, we produced a second one where we assumed a uniform distribution of sources with $S_\nu = 3\sigma_{rms}$.
    Since these sources are exactly at the detection limit, we are not able to properly subtract them; therefore, we do not follow the subtraction procedure for this run and directly investigate their contribution to the profile of the RH.
    Using the LoTSS-DR3 \citep{Shimwell2026} source catalogue, we determined the number density of sources in circular regions of radius $r_{500}$ of the cluster close to A2244 and compared it to the number density of sources within $r_{500}$ of A2244.
    We found an average density of sources of $0.26~\mathrm{sources/arcmin^2}$ both in A2244 and in the nearby regions.
    Therefore, we assumed a uniform distribution of sources for the second simulation, where the density of sources was derived from the MeerKAT L-band image, which resulted in $0.42~\mathrm{sources/arcmin^2}$.
    We refer to this simulation as `Case 2' in the following sections.
    We produced this run to investigate how a large amount of un-subtracted sources can contaminate low surface brightness emission. \par
    We carried out the procedure using the \texttt{Injector of Point SoUrces in Mock observations} (\texttt{IPSUM}) pipeline\footnote{\url{https://github.com/matteocianf/IPSUM}}, presented in this work for the first time. 
    Alongside with LoSiTo, the pipeline uses \texttt{DP3} to work on the visibilities and \textsc{WSClean} for imaging.
    {\renewcommand{\arraystretch}{1.15}
    \begin{table}[!htb]
        \centering
        \caption{Injected and recovered values of the two mock observations.}
        \begin{tabular}{cccccc}
            \hline \hline
            \multirow{2}{*}{Simulation run} & Injected $I_0$ & Injected $r_e$ & \multirow{2}{*}{Mask} & Recovered $I_0$ & Recovered $r_e$ \\
             & $\mathrm{[\mu Jy/arcsec^2]}$ & [kpc] &  & $\mathrm{[\mu Jy/arcsec^2]}$ & [kpc] \\
            \hline
            \multirow{2}{*}{Case 1} & \multirow{2}{*}{$2.7$} & \multirow{2}{*}{$110$} & No & $2.67 \pm 0.21$ & $110 \pm 4$ \\ 
            &  &  & Yes & $2.74 \pm 0.22$ & $108 \pm 4$ \\ 
            \hline 
            Case 2 & $2.7$ & $110$ & - & $2.78 \pm 0.23$ & $103 \pm 4$ \\ 
            \hline
        \end{tabular}
        \label{tab:sim_table}
    \end{table}
    } \par
\begin{figure}[!htb]
        \centering
        \includegraphics[width = 0.34\textwidth]{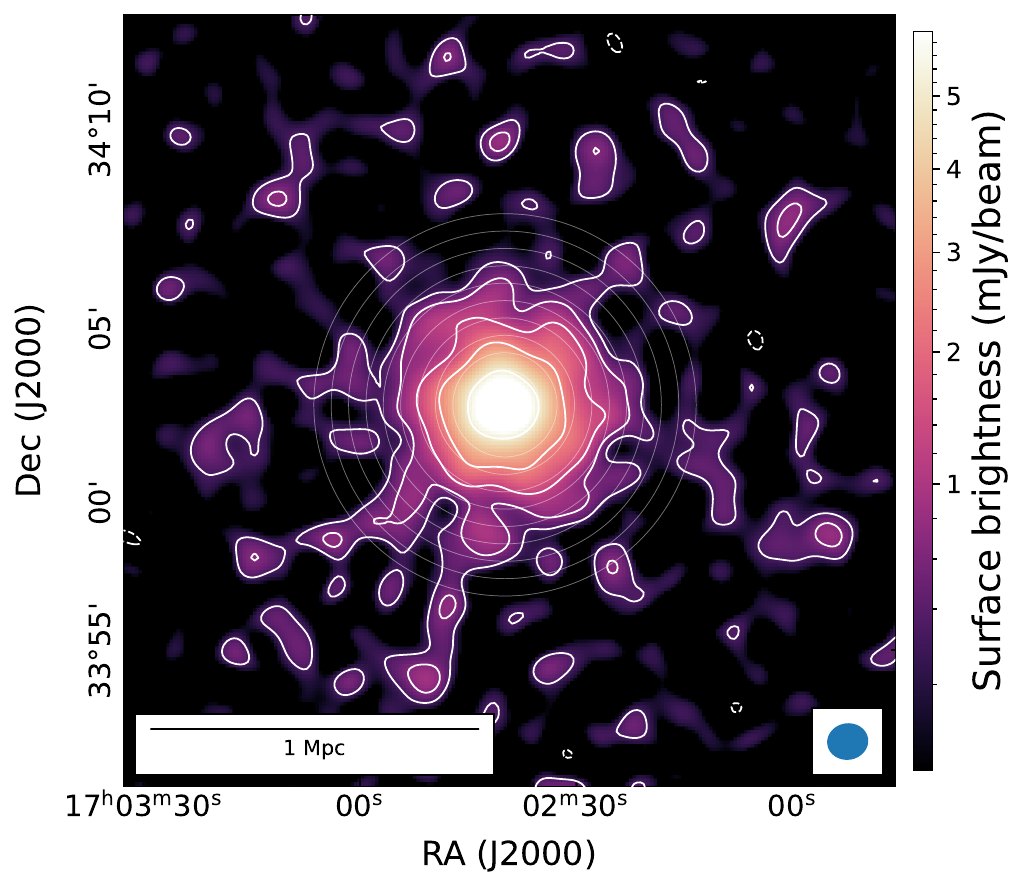}
        \includegraphics[width = 0.30\textwidth]{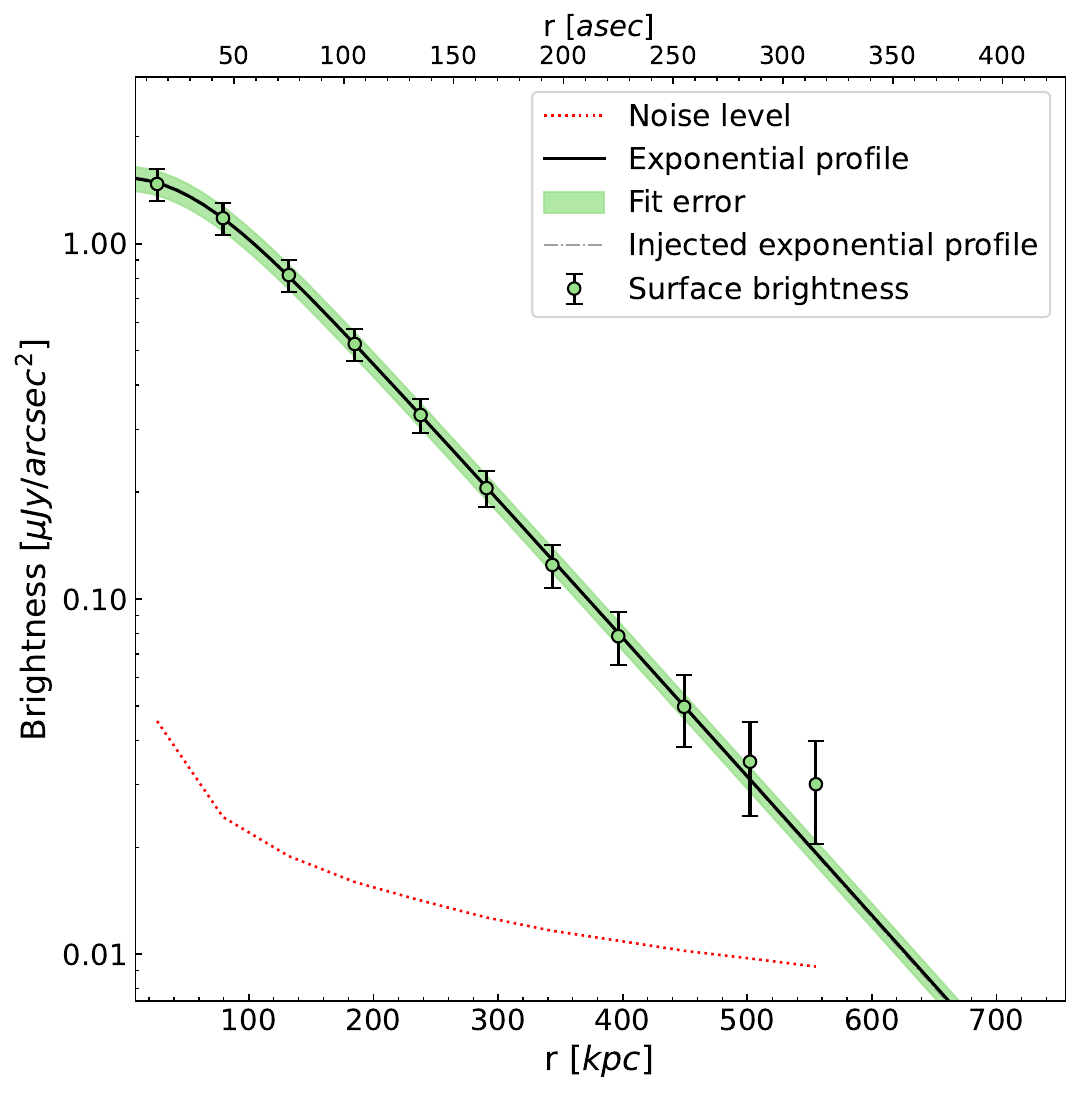}
        \caption{\textit{Left}: Source-subtracted low-resolution ($6\arcsec$) image of the simulated observation. The contours are set at $[-3, 2, 4, 8, 16, 32]\times \sigma_{rms}$, where $\sigma_{rms} = 160~\mathrm{\mu Jy/beam}$. The beam size is shown in the bottom right corner of the image. \textit{Right}: Surface brightness profile.}
        \label{fig:real_nomask}
    \end{figure}
\begin{figure}[!htb]
    \centering
    \includegraphics[width = 0.34\textwidth]{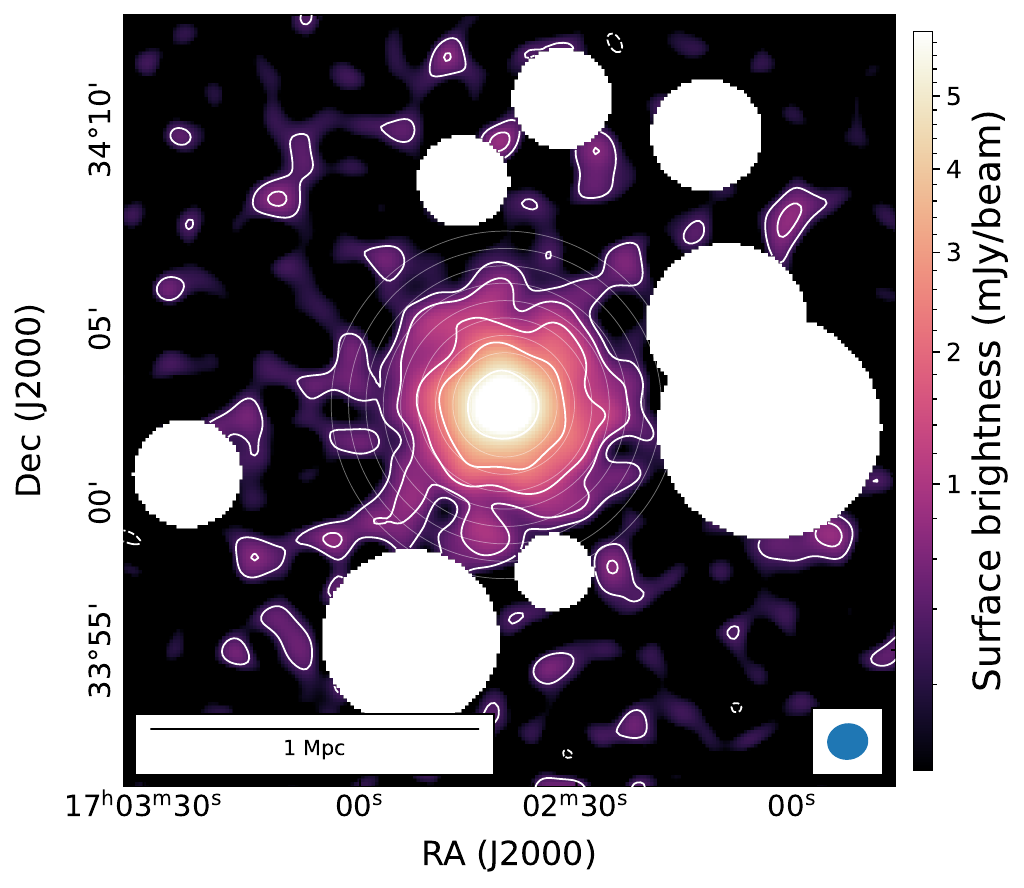}
    \includegraphics[width = 0.32\textwidth]{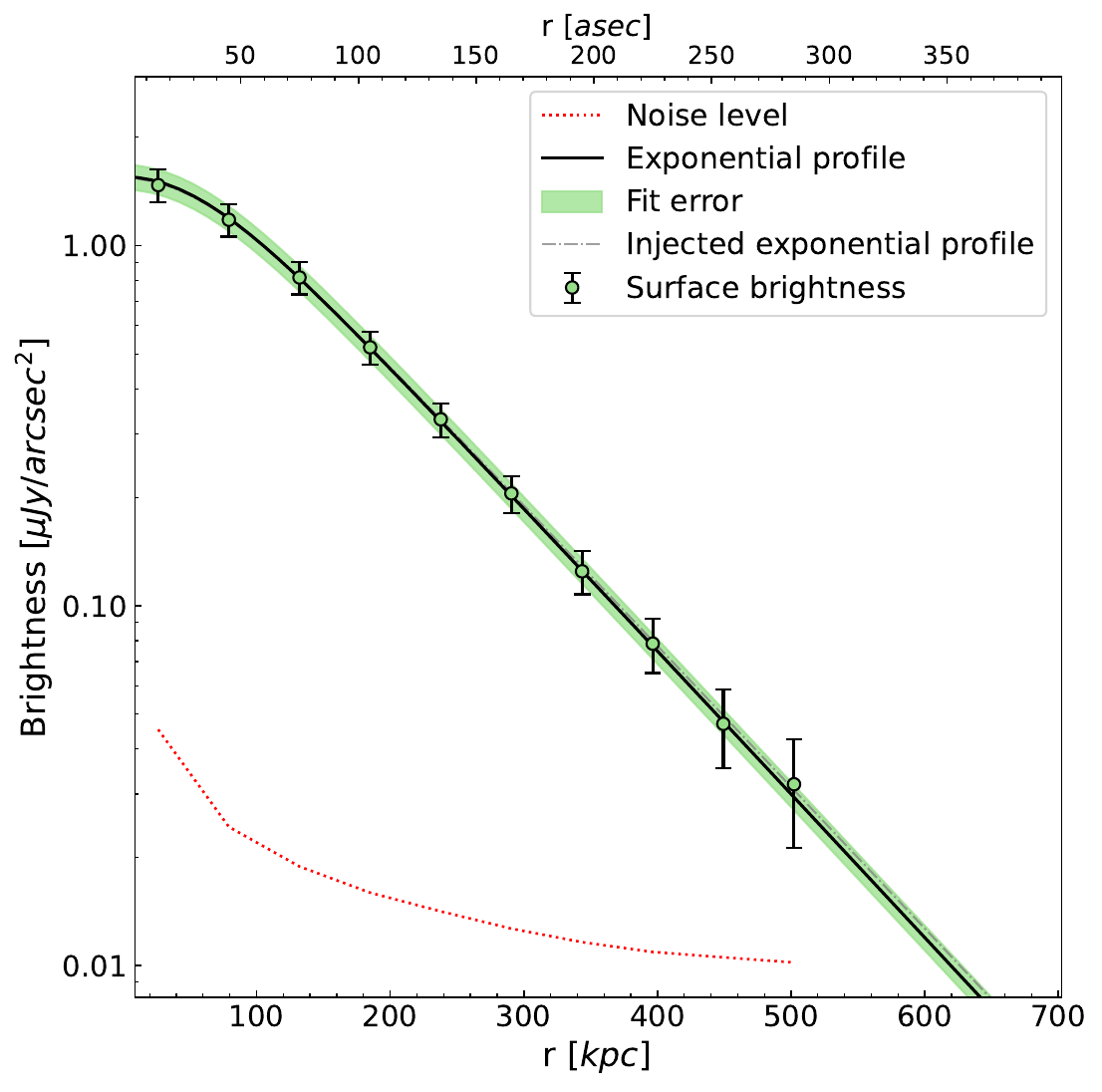}
    \includegraphics[width = 0.34\textwidth]{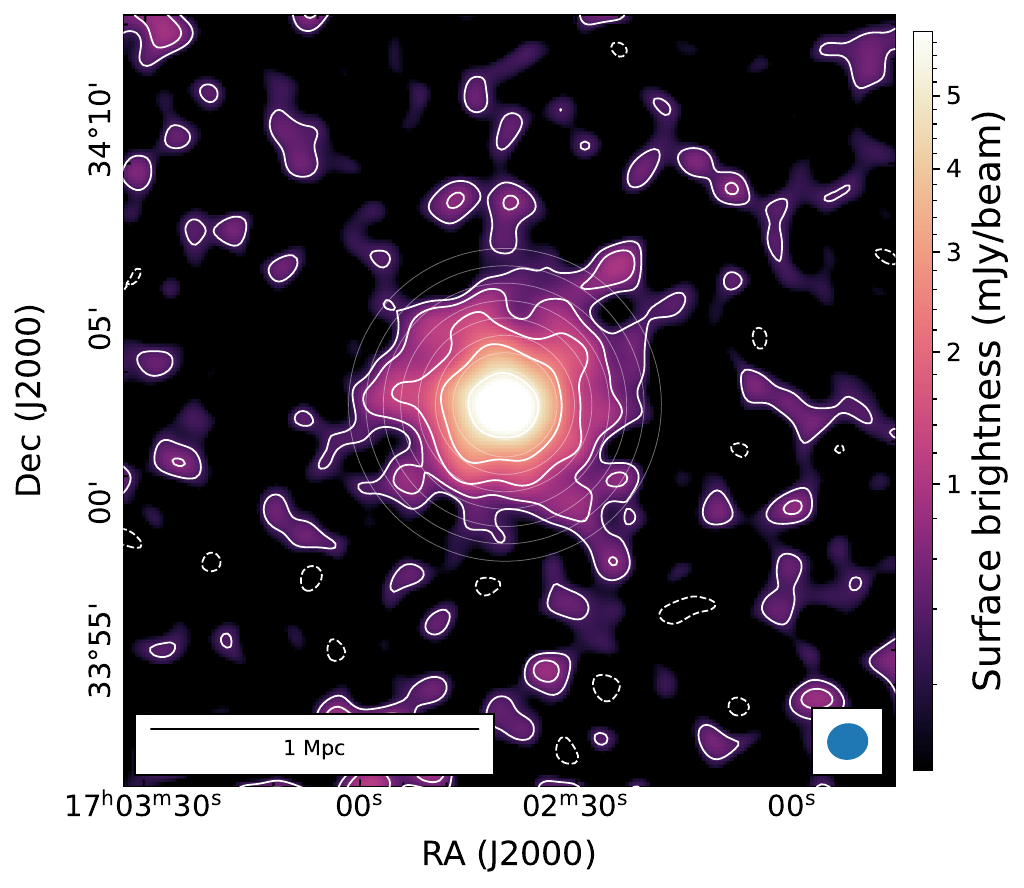}
    \includegraphics[width = 0.32\textwidth]{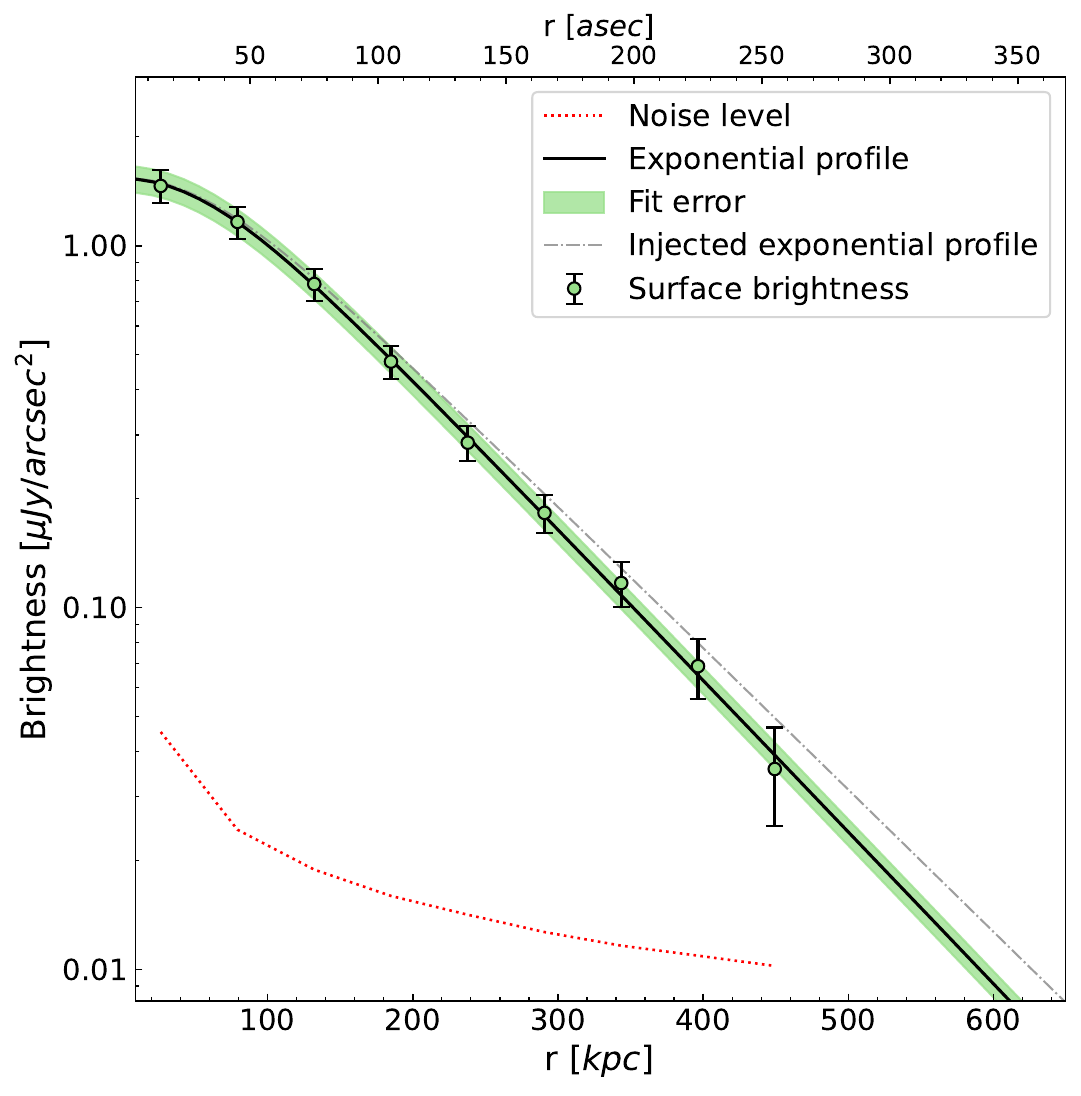}
    \caption{\textit{Left}: Low-resolution ($60\arcsec$) images. The contours are set at $[-3, 2, 4, 8, 16, 32]\times \sigma_{rms}$. The white pixels represent masked regions due to the presence of subtraction residuals. \textit{Right}: Surface brightness profiles. \textit{Top row}: Case 1. \textit{Bottom row}: Case 2.}
    \label{fig:real_mask}
\end{figure}
\begin{figure}[!htb]
    \centering
    \includegraphics[width=0.4\linewidth]{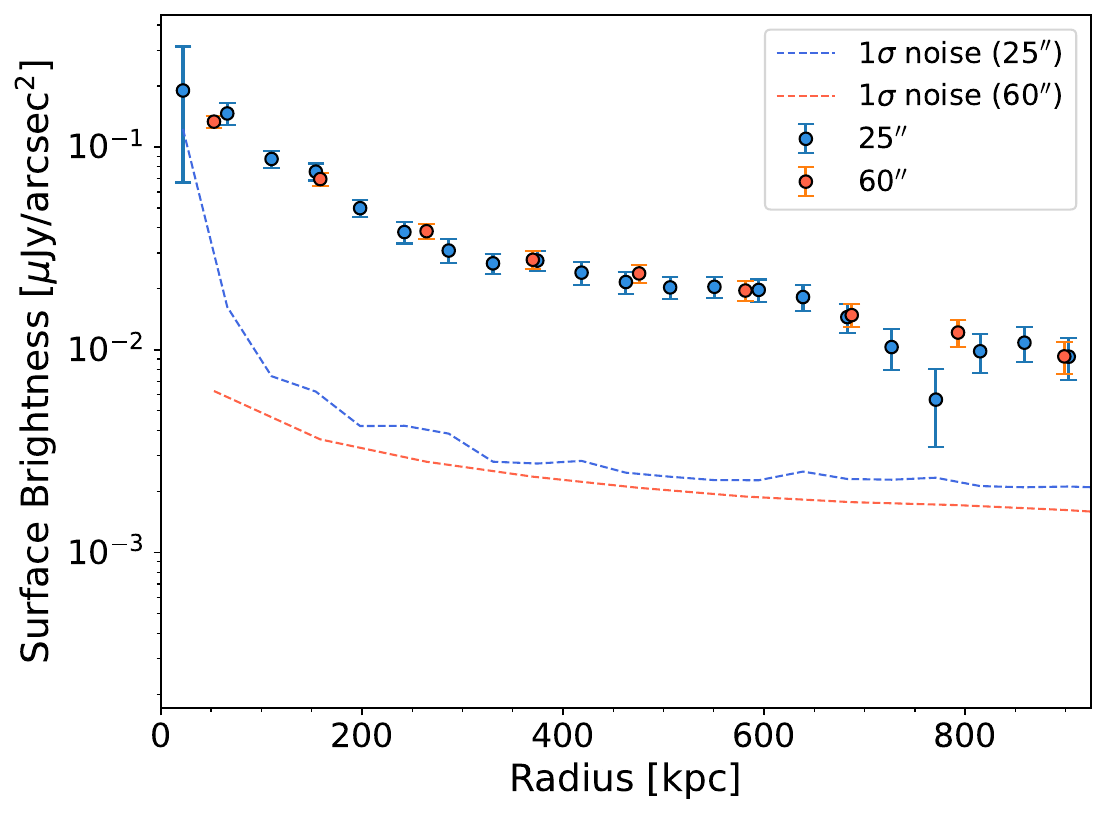}
    \caption{Comparison of the surface brightness profiles in L-band of the diffuse emission at intermediate ($\approx 25\arcsec$, blue) and low ($\approx 60\arcsec$, red) resolution. The dashed lines show the $1\sigma$ detection limit in each annulus for the images.}
    \label{fig:sb_comparison}
\end{figure}

    In the first run, after the subtraction of the sources, we find that the radial profile of the diffuse emission can be fitted by a single exponential profile, with values consistent, within the errors, with the ones of the injected RH, even when we do not mask the residuals.
    The low-resolution image and surface brightness profile are shown in Fig. \ref{fig:real_nomask}.
    From the radial profile, we can see that there is only a small deviation from the exponential model, although it is not significative enough to be considered as second component.
    The best-fit values are $I_0 = 2.67 \pm 0.21 \ \mathrm{\mu Jy/arcsec^2}$ and $r_e = 110 \pm 4 \ \mathrm{kpc}$.
    When we mask the visible subtraction residuals (see top row in Fig. \ref{fig:real_mask}), we retrieve a perfectly exponential profile with parameters $I_0 = 2.74 \pm 0.22 \ \mathrm{\mu Jy/arcsec^2}$ and $r_e = 108 \pm 4 \ \mathrm{kpc}$, which is consistent within the errors with the injected profile and with the results when no masks were applied to the image. 
    These residuals, at low resolutions, give a larger contribution to the radial profile of the diffuse emission, showing as a small deviation from the injected exponential. \par
    As already explained, we decided to test a second case in which all the injected sources have flux density $S_\nu = 3\sigma_{rms}$ and are uniformly distributed in the image.
    From the low-resolution image (bottom row, Fig. \ref{fig:real_mask}) we can see how the sources are weakly contaminating the morphology of the RH, but not the surface brightness profile, which is consistent with the inject exponential model.
    The best-fit parameters for the recovered profile are $I_0 = 2.78 \pm 0.23$ and $r_e = 103 \pm 4~\mathrm{kpc}$.
    The best-fit parameters of the two runs are shown in Tab. \ref{tab:sim_table}. \par
    From these tests, we see that the contribution of faint point sources with flux densities of $S_\nu = 3 \sigma_{rms}$ is negligible and cannot form a outer component around the RH profile.
    The results from the first injection, with the distribution of sources taken from the MeerKAT L-band image, show that a small contribution from sources can actually contaminate the surface brightness profile, only in annuli in which the average brightness is comparable to the one of the RH.
    This contamination, though, is almost negligible, as it only impacts the last point of the profile and with a deviation that is almost consistent within the errors of the fit.
    This contamination might be related to the presence of faint sources in the vicinity of brighter sources, which reduces the ability to subtract point sources, which then leads to the presence of subtraction residuals.
    These residuals, when creating the low-resolution image, are spread over a larger area and can contribute to the flux density of the diffuse emission, increasing the surface brightness with respect to the expected one.
    These residuals appear as patchy emission around the RH and not as diffuse radio emission, making them easy to identify and mask. \par
    These mock observations failed at reproducing the outer component in the RH profile that we observe in A2244, confirming the presence of this large-scale diffuse radio emission in the cluster and the fact that it is not related to faint point sources blending together at low resolutions.
    In general, galaxy clusters host various extended discrete sources and A2244 can be considered a peculiar case, as it only hosts a single faint tailed radio galaxy (T1).
    These extended sources should be injected in more realistic simulations, as they are more difficult to subtract and can leave stronger subtraction residuals.
    Extended radio galaxies can also be related to the origin of megaHs, as they inject CRe in the cluster, which can diffuse and be re-accelerated after cluster or group interactions, although in this case we do not seem to observe any relation between T1 and the megaH.

\section{Surface brightness profiles comparison}
In this section we show the comparison between the surface brightness profiles at low ($60\arcsec$) and mid ($25\arcsec$) resolution in L-band obtained from the source-subtracted images.
\FloatBarrier

\section{Point-to-point correlation grids}
\begin{figure}[!htb]
    \centering
    \includegraphics[width = 0.36\textwidth]{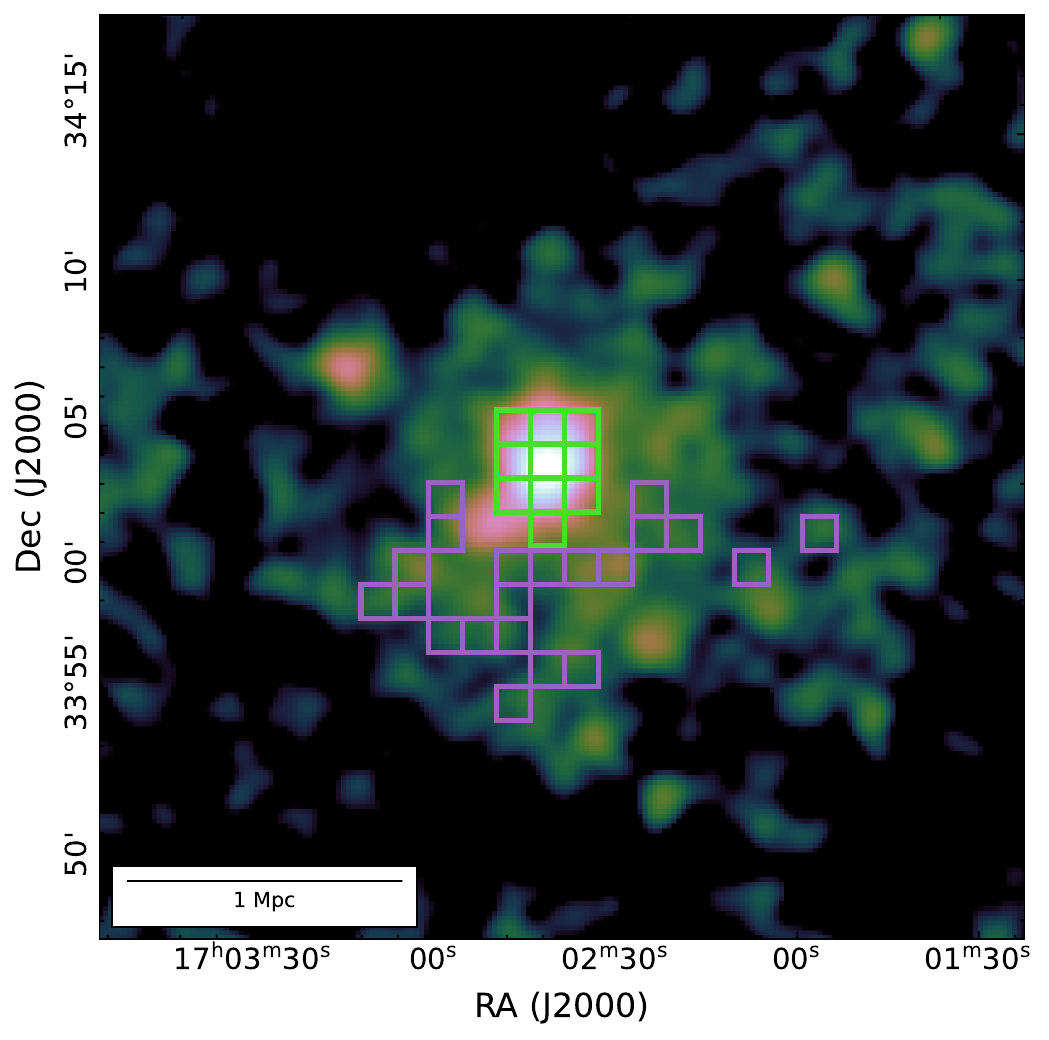}
    \includegraphics[width = 0.36\textwidth]{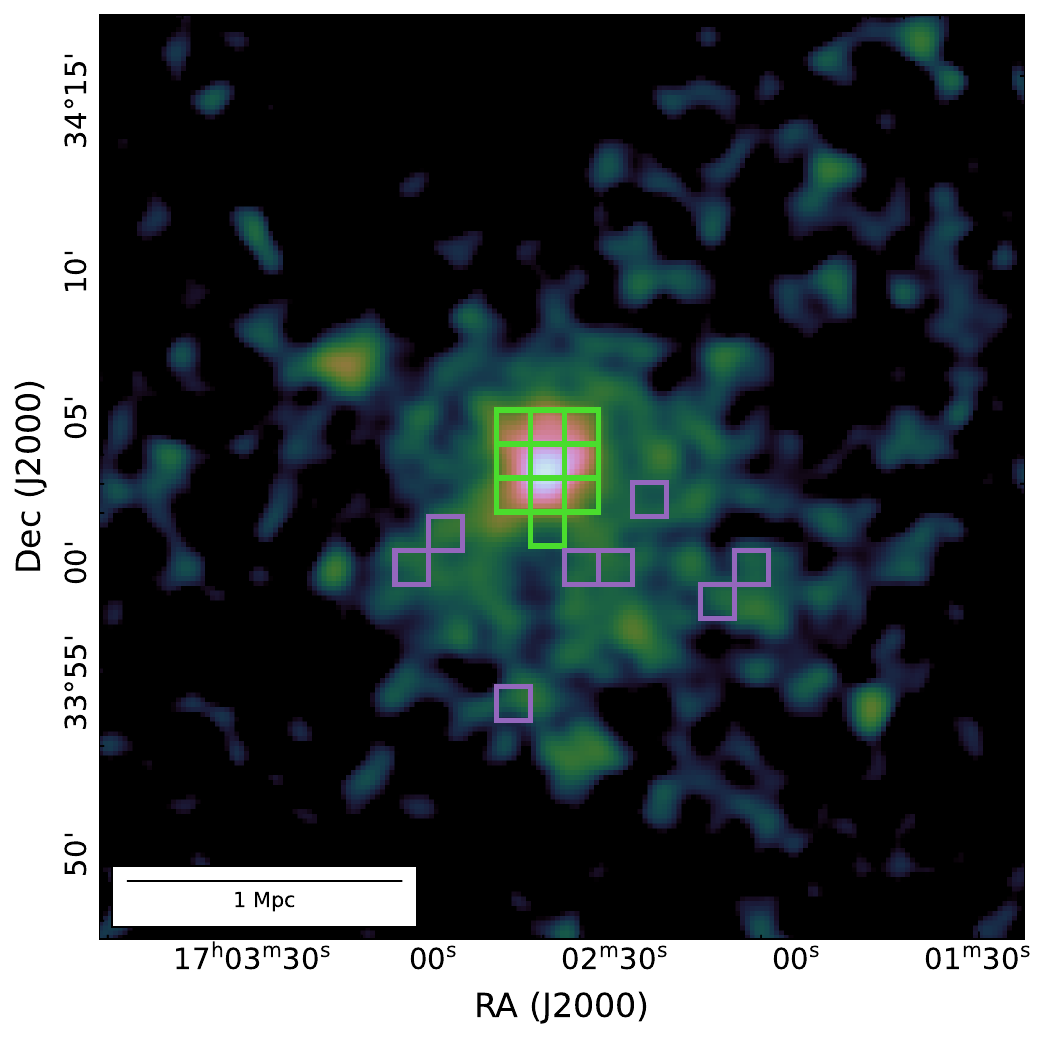}
    \caption{Low-resolution ($\approx 60\arcsec$) images of A2244 at $813~\mathrm{MHz}$ (\textit{left}) and $1279~\mathrm{MHz}$ (\textit{right}) with the cells used to derive the point-to-point radio X-ray correlations. The cells are colour coded in the same way as the point in Figs. \ref{fig:ptp-corr} and \ref{fig:ptp-lband}.}
    \label{appendix_fig:grids}
\end{figure}

In this section we show the grids used to compute the point-to-point correlations at $813~\mathrm{MHz}$ and $1279~\mathrm{MHz}$.
We generated a grid of cells within $r_{500}$ of the cluster, while carefully masking the regions in which radio residuals or X-ray sources not associated with the cluster are present. The cells shown in Fig. \ref{appendix_fig:grids} are the ones that have a surface brightness greater than $2\sigma_{rms}$ and that have been used for the point-to-point correlation. The size of the cells is $70\arcsec \times 70\arcsec$.
\FloatBarrier

\end{appendix}
\end{document}